\documentclass[a4paper]{article}
\usepackage{arxiv}
\pdfoutput=1

\usepackage[T1]{fontenc}
\usepackage[utf8]{inputenc}
\usepackage{balance}

\usepackage[binary-units]{siunitx}
\DeclareSIUnit\Byte{\text{Byte}}
\usepackage{etoolbox}
\robustify\bfseries

\usepackage[inline]{enumitem}
\setlist{itemsep=2pt,parsep=0pt,topsep=2pt}
\setlist[1]{labelindent=\parindent}

\usepackage{algorithm}
\usepackage[noend]{algpseudocode}

\usepackage{pgfplotstable}
\usepackage{colortbl}
\usepackage{tipa}

\usepackage{tikz}
\usetikzlibrary{shapes.arrows,chains,calc}
\usetikzlibrary{decorations.pathreplacing,matrix}

\usepackage{rotating}
\usepackage{pdflscape}
\usepackage{afterpage}
\usepackage[labelformat=parens]{subcaption}

\usepackage{amsthm}%
\newtheorem{theorem}{Theorem}[section]%

\usepackage{amsfonts}

\newcommand{\realrange}[2]{\left[#1, #2\right]}

\newcommand{\unitrange}[2]{\realrange{0}{1}}

\newcommand{\discussionsize}{\small}

\newcommand{\frage}[1]{}

\newsavebox{\codeparam}

{\end{disscodepos}}

\newdimen\endofsize\endofsize=0.5em
\def\endofbeweis{~\quad\hglue\hsize minus\hsize
                 \hbox{\vrule height \endofsize width
\endofsize}\par}

\usepackage{relsize}

\newcommand{\ie}{i.e.}
\newcommand{\eg}{e.g.}
\newcommand{\ea}{et\ al.}
\newcommand{\bigO}{\mathcal{O}}

\newcommand{\Query}{\textsc{Query}}

\newcommand{\InDegree}[1]{\ensuremath{\mathrm{deg}^\mathsmaller{-}(#1)}}
\newcommand{\OutDegree}[1]{\ensuremath{\mathrm{deg}^\mathsmaller{+}(#1)}}
\newcommand{\True}{\texttt{true}}
\newcommand{\False}{\texttt{false}}

\newcommand{\ReachableSet}{\textsf{R}}

\newcommand{\Card}[1]{\ensuremath{|#1|}}
\newcommand{\List}[1]{\ensuremath{\mathcal{L}}}
\newcommand{\InNeighborList}[1]{\ensuremath{\mathcal{N}^-}}
\newcommand{\OutNeighborList}[1]{\ensuremath{\mathcal{N}^+}}
\newcommand{\InNeighborHood}[1]{\ensuremath{\textsf{N}^{\mathsmaller{-}}}(#1)}
\newcommand{\OutNeighborHood}[1]{\ensuremath{\textsf{N}^{\mathsmaller{+}}}(#1)}
\newcommand{\InEdgesList}[1]{\ensuremath{\mathcal{E}^-}}
\newcommand{\OutEdgesList}[1]{\ensuremath{\mathcal{E}^+}}

\newcommand{\Queue}[1]{\Code{Q}}

\newcommand{\Range}[1]{\left[#1\right]}

\newcommand{\PathTo}{\ensuremath{\to^*}}
\newcommand{\NoPathTo}{\ensuremath{\not\to^*}}
\newcommand{\OutReachability}[1]{\ReachableSet^\mathsmaller{+}(#1)}
\newcommand{\InReachability}[1]{\ReachableSet^\mathsmaller{-}(#1)}
\newcommand{\High}[1]{#1_H}
\newcommand{\Low}[1]{#1_L}
\newcommand{\Max}[1]{#1_{X}}
\newcommand{\Min}[1]{#1_{N}}

\newcommand{\SCC}{\text{SCC}}
\newcommand{\SCCof}{\mathcal{S}}
\newcommand{\CC}{\text{WCC}}

\newcommand{\DAG}{DAG}
\newcommand{\TopSort}{\tau}
\newcommand{\Reverse}[1]{{#1}^{\mathsmaller{\mathrm{R}}}}
\newcommand{\Condensed}[1]{{#1}^{\mathsmaller{\mathrm{C}}}}
\newcommand{\TopSortLevel}{L}
\newcommand{\TopSortLevelForward}{\mathcal{F}}
\newcommand{\TopSortLevelBackward}{\mathcal{B}}
\newcommand{\ParamNumSupports}{k}
\newcommand{\ParamSupportSampleSize}{p}
\newcommand{\ParamNumTopsorts}{t}
\newcommand{\ParamLevelThreshold}{h}

\newcommand{\OutReachSet}{{R^\mathsmaller{+}}}
\newcommand{\InReachSet}{{R^\mathsmaller{-}}}
\newcommand{\AnswerableNegQueries}{\NegQueries}
\newcommand{\AnswerablePosQueries}{\PosQueries}
\newcommand{\PosQueries}{\mathcal{P}}
\newcommand{\NegQueries}{\mathcal{N}}
\newcommand{\ReachabilityRatio}{\rho}

\newcommand{\AnswerableNegQueryRatio}{\ReachabilityRatio^{\mathsmaller{-}}}
\newcommand{\AnswerablePosQueryRatio}{\ReachabilityRatio^{\mathsmaller{+}}}
\newcommand{\Sources}{\mathcal{S}}
\newcommand{\Sinks}{\mathcal{T}}
\newcommand{\Isolated}{\mathcal{I}}

\newcommand{\Reachability}{\textsl{Reachability}}

\newcommand{\Code}[1]{\texttt{#1}}

\newcommand{\Rplus}{\protect\hspace{-.1em}\protect\raisebox{.35ex}{\smaller{\smaller\textbf{+}}}}
\newcommand{\Cpp}{\mbox{C\Rplus\Rplus}\xspace}
\newcommand{\Tool}[1]{\textsf{#1}}

\newcommand{\SNAP}{\Tool{SNAP}}

\newcommand{\Preach}{\texttt{PReaCH}}

\newcommand{\Grail}{\texttt{GRAIL}}
\newcommand{\Sreach}{\texttt{O'Reach}}
\newcommand{\SrPlus}{\texttt{O'R\,+}}
\newcommand{\FullMatrix}{\texttt{Matrix}}
\newcommand{\TF}{\texttt{TF}}
\newcommand{\PPL}{\texttt{PPL}}
\newcommand{\Instance}[1]{\textsl{#1}}
\newcommand{\pBFS}{\texttt{pBiBFS}}
\newcommand{\IP}{\texttt{IP}}
\newcommand{\BFL}{\texttt{BFL}}
\newcommand{\IPsparse}{\texttt{IP(s)}}
\newcommand{\IPdense}{\texttt{IP(d)}}
\newcommand{\BFLsparse}{\texttt{BFL(s)}}
\newcommand{\BFLdense}{\texttt{BFL(d)}}

\newcommand{\kIP}{\ensuremath{k_{\IP}}}
\newcommand{\hIP}{\ensuremath{h_{\IP}}}
\newcommand{\dBFL}{\ensuremath{d_{\BFL}}}
\newcommand{\sBFL}{\ensuremath{s_{\BFL}}}

\newcommand{\TabLabel}[1]{\label{tab:#1}}
\newcommand{\Table}[1]{Table~\ref{tab:#1}}

\newcommand{\Figure}[1]{Figure~\ref{fig:#1}}

\newcommand{\SectLabel}[1]{\label{sect:#1}}
\newcommand{\Section}[1]{Sect.~\ref{sect:#1}}
\newcommand{\ObLabel}[1]{\label{ob:#1}}
\newcommand{\Observation}[1]{Observation~\ref{ob:#1}}
\newcommand{\Ob}[1]{\ref{ob:#1}}
\newcommand{\ItemLabel}[1]{\label{itm:#1}}
\newcommand{\Item}[1]{\ref{itm:#1}}
\newcommand{\ThmLabel}[1]{\label{thm:#1}}
\newcommand{\Theorem}[1]{Theorem~\ref{thm:#1}}

\newcommand{\Algorithm}[1]{Figure~\ref{alg:#1}}

\newcommand{\FootLabel}[1]{\label{foot:#1}}

\newcommand{\FastTrueAnswer}{\tikz\node[circle,draw=green!50!black,thick,inner sep=2pt]{};}
\newcommand{\FastFalseAnswer}{\tikz\node[circle,draw=red!50!black,thick,fill=red!50!black,inner sep=2pt]{};}

\usepackage{amsmath}
\hypersetup{%
  colorlinks,%
  linkcolor={red!50!black},%
  citecolor={green!50!black},%
  urlcolor={blue!50!black}%
}%

\usepackage{xspace}
\usepackage{graphics}
\usepackage{booktabs}

\newcommand{\etal}{et~al.\xspace}

\newcommand{\wrt}{w.\,r.\,t.\xspace}

\newbool{longversion}
\booltrue{longversion}%

\title{\Sreach{}:\\Even Faster Reachability in Large Graphs}
\titlerunning{\Sreach{}: Even Faster Reachability in Large Graphs}
\author{Kathrin Hanauer}{University of Vienna, Faculty of Computer Science, Vienna, Austria}{kathrin.hanauer@univie.ac.at}{https://orcid.org/0000-0002-5945-837X}{}

\author{Christian Schulz}{Heidelberg University, Heidelberg, Germany}{christian.schulz@informatik.uni-heidelberg.de}{https://orcid.org/0000-0002-2823-3506}{}

\author{Jonathan Trummer}{University of Vienna, Faculty of Computer Science, Vienna, Austria}{jonathan.trummer@univie.ac.at}{https://orcid.org/0000-0002-1086-4756}{}
\authorrunning{K.\ Hanauer, C.\ Schulz, J.\ Trummer}
\funding{The research leading to these results has received funding %
from the European Research Council under the European Union’s %
Seventh Framework Programme (FP/2007-2013) / ERC Grant %
Agreement no.\ 340506.}%
\date{}%
\begin{document}

\maketitle
\begin{abstract}
One of the most fundamental problems in computer science is the \emph{reachability
problem}: Given a directed graph and two vertices $s$ and $t$, can $s$
\emph{reach} $t$ via a path?
We revisit existing techniques and combine them with new approaches to support
a large portion of \emph{reachability queries} in constant time using a
linear-sized \emph{reachability index}.
Our new algorithm \Sreach{} can be easily combined with previously developed
solutions for the problem or run standalone.

In a detailed experimental study, we compare a variety of algorithms with
respect to their index-building and query times as well as their memory
footprint on a diverse set of instances.
Our experiments indicate that the query performance often depends strongly not
only on the type of graph, but also on the result, i.e., \emph{reachable} or
\emph{unreachable}.
Furthermore, we show that previous algorithms are significantly sped up
when combined with our new approach in almost all scenarios. %
Surprisingly, due to cache effects, a higher investment in space doesn't
necessarily pay off:
\emph{Reachability queries} can often be answered even faster than single
memory accesses in a precomputed full reachability matrix.
\end{abstract}

\section{Introduction}
Graphs are used to model problem settings of various different disciplines.
A natural question that arises frequently is whether one vertex of the graph
can \emph{reach} another vertex via a path of directed edges.
\emph{Reachability} finds application in a wide variety of fields, such as program
and dataflow analysis~\cite{reps1998program, reps1995precise}, user-input
dependence analysis~\cite{scholz2008userinput}, XML query
processing~\cite{DBLP:conf/icde/WangHYYY06}, and more~\cite{Yu2010}.
Another prominent example is the Semantic Web which is composed of RDF/OWL data. These are often very huge graphs with rich content. Here, reachability queries are often necessary to deduce relationships among the objects.

There are two straightforward solutions to the reachability problem:
The first is to answer each query individually with a graph traversal algorithm,
such as breadth-first search (BFS) or depth-first search (DFS), in worst-case
$\bigO(m+n)$ time and $\bigO(n)$ space.
Secondly, we can precompute a full all-pairs reachability matrix in an
initialization step and answer all ensuing queries in worst-case constant time.
In return, this approach suffers from a space complexity of $\bigO(n^2)$ and an
initialization time of $\bigO(n\cdot m)$ using the Floyd-Warshall
algorithm~\cite{Floyd62,Warshall62,CLRS09} or starting a graph traversal at
each vertex in turn.
Alternatively, the initialization step can be performed in $\bigO(n^\omega)$
via fast matrix multiplication, where $\bigO(n^\omega)$ is the time required to
multiply two $n \times n$ matrices ($2 \leq \omega < 2.38$~\cite{LeGall14}).
With increasing graph size, however, both the initialization time and space
complexity of this approach become impractical.
We therefore strive for alternative algorithms which decrease these
complexities whilst still providing fast query lookups.

\textbf{Contribution.}
In this paper, we study a variety of approaches that are able to support fast
\emph{reachability queries}.
All of these algorithms perform some kind of preprocessing on the graph and
then use the collected data to answer reachability queries in a timely manner.
Based on simple observations, we provide a new algorithm, \Sreach{}, that can
improve the query time for a wide range of cases over state-of-the-art
reachability algorithms at the expense of some additional precomputation time
and space or be run standalone.
Furthermore, we show that previous algorithms are significantly sped up
when combined with our new approach in almost all scenarios.
In addition, we show that the expected query performance of various
algorithms does not only depend on the type of graph, but also on the ratio of
successful queries, i.e., with result \emph{reachable}.
Surprisingly, through cache effects and a significantly smaller memory
footprint, especially unsuccessful \emph{reachability queries} can be answered
faster than single memory accesses in a precomputed reachability matrix.

\section{Preliminaries}\SectLabel{preliminaries}
\textbf{Terms and Definitions.}
Let $G=(V, E)$ be a simple directed graph with vertex set $V$
and edge set $E\subseteq V \times V$.
As usual, $n=|V|$ and $m=|E|$.
An edge $(u, v)$ is said to be \emph{outgoing} at $u$ and \emph{incoming}
at $v$, and $u$ and $v$ are called \emph{adjacent}.
The \emph{out-degree} $\OutDegree{u}$ (\emph{in-degree} $\InDegree{u}$) of a
vertex $u$ is its number of outgoing (incoming) edges.
A vertex without incoming (outgoing) edges is called a
\emph{source} (\emph{sink}).
The \emph{out-neighborhood} $\OutNeighborHood{v}$
(\emph{in-neighborhood} $\InNeighborHood{v}$) of a vertex $u$ is the set
of all vertices $v$ such that $(u, v) \in E$ ($(v, u) \in E$).
The \emph{reverse} of an edge $(u, v)$ is an edge $(v, u) = \Reverse{(u, v)}$.
The \emph{reverse} $\Reverse{G}$ of a graph $G$ is obtained by keeping the
vertices of $G$, but substituting each edge $(u, v) \in E$ by its reverse, \ie,
$\Reverse{G} = (V, \Reverse{E})$.

A sequence of vertices $s = v_0 \to \dots \to v_k = t$, $k \geq 0$, such that
for each pair of consecutive vertices $v_i \to v_{i+1}$, $(v_i, v_{i+1})\in E$,
is called an \emph{s-t path}.
If such a path exists, $s$ is said to \emph{reach} $t$ and we write $s \PathTo
t$ for short, and $s \NoPathTo t$ otherwise.
The \emph{out-reachability} $\OutReachability{u} = \{ v \mid u \PathTo v\}$
(\emph{in-reachability} $\InReachability{u} = \{v \mid v \PathTo u\}$) of a
vertex $u \in V$ is the set of all vertices that $u$ can reach (that can reach
$u$).

A \emph{weakly connected component (\CC)} of $G$ is a maximal set of vertices $C
\subseteq V$ such that $\forall u, v \in C: u \PathTo v$ in $G=(V, E \cup
\Reverse{E})$, \ie, also using the reverse of edges.
Note that if two vertices $u, v$ reside in different \CC{}s, then $u \NoPathTo
v$ and $v \NoPathTo u$.
A \emph{strongly connected component (\SCC{})} of $G$ denotes a maximal set of
vertices $S \subseteq V$ such that $\forall u, v \in S: u \PathTo v \wedge
v \PathTo u$ in $G$.
Contracting each \SCC{} $S$ of $G$ to a single vertex $v_S$, called its
\emph{representative}, while preserving edges between different \SCC{}s as
edges between their corresponding representatives, yields the
\emph{condensation} $\Condensed{G}$ of $G$.
We denote the \SCC{} a vertex $v \in V$ belongs to by $\SCCof(v)$.
A directed graph $G$ is \emph{strongly connected} if it only has a single
\SCC{} and \emph{acyclic} if each \SCC{} is a singleton, \ie, if $G$ has $n$
\SCC{}s.
Observe that $G$ and $\Reverse{G}$ have exactly the same \CC{}s and \SCC{}s
and that $\Condensed{G}$ is a directed acyclic graph (\DAG{}).
Weakly connected components of a graph can be computed in $\bigO(n+m)$ time,
\eg, via a breadth-first search that ignores edge directions.
The strongly connected components of a graph can be computed in linear
time~\cite{Tarjan72} as well.

A \emph{topological ordering} $\TopSort: V \rightarrow \mathbb{N}_0$ of a
\DAG{} $G$ is a total ordering of its vertices such
that $\forall (u, v) \in E: \TopSort(u) < \TopSort(v)$.
Note that the topological ordering of $G$ isn't necessarily unique, \ie, there
can be multiple different topological orderings.
For a vertex $u \in V$, the \emph{forward topological level} $\TopSortLevelForward(u) =
\min_\TopSort \TopSort(u)$, \ie, the minimum value of $\TopSort(u)$ among all
topological orderings $\TopSort$ of $G$.
Consequently, $\TopSortLevelForward(u) = 0$  if and only if $u$ is a source.
The \emph{backward topological level} $\TopSortLevelBackward(u)$ of $u \in V$
is the topological level of $u$ with respect to $\Reverse{G}$ and
$\TopSortLevelBackward(u) = 0$ if and only if $u$ is a sink.
A topological ordering as well as the forward and backward topological levels
can be computed in linear time~\cite{Kahn62,tarjan1976edge,CLRS09}, see also
\Section{algorithms}.

A \emph{reachability query} \Query($s, t$) for a pair of vertices $s, t \in V$
is called \emph{positive} and answered with \True{} if $s \PathTo t$, and
otherwise \emph{negative} and answered with \False{}.
Trivially, \Query($v, v$) is always \True{}, which is why we only consider
\emph{non-trivial} queries between distinct vertices
$s \neq t \in V$
from here on. %
Let
$\PosQueries$ ($\NegQueries$) denote the set of all positive (negative)
non-trivial queries of $G$, \ie, the set of all $(s, t) \in V \times V$, $s \neq
t$, such that \Query($s, t$) is positive (negative).
The \emph{reachability} $\ReachabilityRatio$ in %
$G$ is the ratio of positive queries among
all non-trivial queries, \ie, $\ReachabilityRatio = \frac{\Card{\PosQueries}}{n(n-1)}$.
Note, that due to the restriction to non-trivial queries%
\footnote{Otherwise, $\frac{1}{n} \leq \ReachabilityRatio$.}%
, $0 \leq \ReachabilityRatio \leq 1$.
The \emph{\Reachability{} problem}, studied in this paper, consists in
answering a sequence of reachability queries for arbitrary pairs
of vertices on a given input graph $G$.

\textbf{Basic Observations.}
With respect to processing a reachability \Query($s, t$) in a graph $G$ for an
arbitrary pair of vertices $s \neq t \in V$, the following basic observations
are immediate and have partially also been noted elsewhere~\cite{merz2014preach}%
:
\begin{enumerate}[label=(B\arabic{*}),series=basic-obs]
\addtolength{\itemindent}{1cm}
\item\ObLabel{basic:sink}\ObLabel{basic:source}
If $s$ is a sink or $t$ is a source, then $s \NoPathTo t$.
\item\ObLabel{basic:cc}
If $s$ and $t$ belong to different \CC{}s of $G$, then $s \NoPathTo t$.
\item\ObLabel{basic:scc}
If $s$ and $t$ belong to the same \SCC{} of $G$, then $s \PathTo t$.
\item\ObLabel{basic:topsort}
If $\TopSort(\SCCof(t)) < \TopSort(\SCCof(s))$ for any topological ordering
$\TopSort$ of  $\Condensed{G}$, then $s \NoPathTo t$.
\end{enumerate}
As mentioned above, the precomputations necessary for
Observations~\Ob{basic:cc} and~\Ob{basic:scc} can be performed in $\bigO(n+m)$
time.
Note, however, that Observations~\Ob{basic:scc} and~\Ob{basic:topsort}
together are \emph{equivalent} to asking whether $s \PathTo t$:
If $s \PathTo t$ and $\SCCof(s) \neq \SCCof(t)$, then
for every topological ordering
$\TopSort$, $\TopSort(\SCCof(s)) < \TopSort(\SCCof(t))$.
Otherwise, if $s \NoPathTo t$, a topological ordering $\TopSort$ with
$\TopSort(\SCCof(t)) < \TopSort(\SCCof(s))$ can be computed by topologically
sorting $\Condensed{G} \cup \{(\SCCof(t),\SCCof(s))\}$.
Hence, the precomputations necessary for \Observation{basic:topsort} would
require solving the \Reachability{} problem for all pairs of vertices already.
Furthermore, a \DAG{} can have exponentially many different topological
orderings.
In consequence, weaker forms are employed, such as the
following~\cite{yildrim2010grail,yildirim2012grail,merz2014preach} (see also
\Section{algorithms}):
\begin{enumerate}[resume*=basic-obs]
\addtolength{\itemindent}{1cm}
\item\ObLabel{basic:ts-level-fw}
If $\TopSortLevelForward(\SCCof(t)) < \TopSortLevelForward(\SCCof(s))$ \wrt
$\Condensed{G}$, then $s \NoPathTo t$.
\item\ObLabel{basic:ts-level-bw}
If $\TopSortLevelBackward(\SCCof(s) < \TopSortLevelBackward(\SCCof(t))$ \wrt
$\Condensed{G}$, then $s \NoPathTo t$.
\end{enumerate}
\textbf{Assumptions.}
Following the convention introduced in preceding
work~\cite{yildrim2010grail,yildirim2012grail,DBLP:conf/sigmod/ChengHWF13,merz2014preach}
(cf.~\Section{related}), we only consider \Reachability{} on \DAG{}s from here
on and implicitly assume that the condensation, if necessary, has already been
computed and \Observation{basic:scc} has been applied.
For better readability, we also drop the use of $\SCCof(\cdot)$.

\section{Related Work}\SectLabel{related}
\begin{table}[h!]
\footnotesize
\setlength{\tabcolsep}{1.5pt}
\caption{Time and space complexity of reachability algorithms.
Parameters: %
$\kIP{}$: \#permutations,
$\hIP{}$: \#vertices with precomputed $\OutReachability{\cdot}$,
$\sBFL{}$: size of Bloom filter (bits),
$\ReachabilityRatio{}$: reachability in $G$,
$\ParamNumTopsorts$: \#topological orderings,
$\ParamNumSupports$: \#supportive vertices,
$\ParamSupportSampleSize$: \#candidates per supportive vertex
}%
\TabLabel{algorithm-complexities}
\centering
\begin{tabular}{@{}lcccc@{}}
\toprule
Algorithm & Initialization Time & Index Size (\si{\Byte}) & Query Time  & Query Space \\
\midrule
BFS/DFS
&
$\mathcal{O}(1)$
&
$0$
&
$\mathcal{O}(n+m)$
&
$\mathcal{O}(n)$
\\
Full matrix
&
$\mathcal{O}(n \cdot (n+m))$
&
$n^2/8$
&
$\mathcal{O}(1)$
&
$\mathcal{O}(1)$
\\
\PPL{}~\cite{DBLP:conf/cikm/YanoAIY13}
&
$\mathcal{O}(n\log n+m)$
&
$\mathcal{O}(n \log n)$
&
$\mathcal{O}(\log n)$
&
$\mathcal{O}(\log n)$
\\
\Preach{}~\cite{merz2014preach}
&
$\mathcal{O}(m+n\log n)$
&
$56n$
&
$\bigO(1)$ or $\bigO(n+m)$
&
$\mathcal{O}(n)$
\\
\IP{}(\kIP{}, \hIP{})~\cite{DBLP:journals/vldb/WeiYLJ18}
&
$\mathcal{O}((\kIP{} + \hIP{})(n+m))$
&
$\mathcal{O}((\kIP{} + \hIP{})n)$
&
$\mathcal{O}(\kIP{})$ or $\mathcal{O}(\kIP{}\cdot n \cdot \ReachabilityRatio{}^2)$
&
$\mathcal{O}(n)$
\\
\BFL{}(\sBFL{})~\cite{DBLP:journals/tkde/SuZWY17}
&
$\mathcal{O}(\sBFL{}\cdot(n+m))$
&
$2\lceil\frac{\sBFL{}}{8}\rceil n$
&
$\mathcal{O}(\sBFL{})$ or $\mathcal{O}(\sBFL{}\cdot n + m)$
&
$\mathcal{O}(n)$
\\
\Sreach{}($\ParamNumTopsorts, \ParamNumSupports, \ParamSupportSampleSize$)~(\Section{algorithms})
&
$\bigO((\ParamNumTopsorts+\ParamNumSupports\ParamSupportSampleSize)(n+m))$
&
$(12 + 12 \ParamNumTopsorts
+ 2 \lceil\frac{\ParamNumSupports}{8}\rceil)n$%
&
$\bigO(\ParamNumSupports + \ParamNumTopsorts + 1)$ or $\bigO(n+m)$
&
$\mathcal{O}(n)$
\\
\bottomrule
\end{tabular}
\end{table}
A large amount of research on reachability indices has been conducted.
Existing approaches can roughly be put into three categories:
compression of transitive closure~\cite{DBLP:conf/sigmod/JinRDY12,%
DBLP:journals/tods/Jagadish90,%
DBLP:conf/icde/ChenC08,%
DBLP:conf/icde/WangHYYY06,%
DBLP:journals/tods/JinRXW11,%
DBLP:conf/sigmod/SchaikM11},
hop-labeling-based algorithms~\cite{DBLP:journals/siamcomp/CohenHKZ03,%
DBLP:conf/edbt/ChengYLWY06,%
DBLP:conf/edbt/SchenkelTW04,%
DBLP:conf/cikm/YanoAIY13,%
DBLP:journals/pvldb/JinW13},
as well as pruned search~\cite{DBLP:conf/sigmod/JinXRW08,%
DBLP:conf/sigmod/TrisslL07,%
yildrim2010grail,yildirim2012grail,merz2014preach,veloso2014reachability,%
DBLP:journals/vldb/WeiYLJ18,%
DBLP:journals/tkde/SuZWY17}.
As Merz and Sanders~\cite{merz2014preach} noted, the first category gives very
good query times for small networks, but doesn't scale very well to large
networks (which is the focus of this work).
Therefore, we do not consider approaches based on
this technique more closely.
Hop labeling algorithms typically build paths from labels that are stored for
each vertex.
For example in 2-hop labeling, each vertex stores two sets containing vertices it
can reach in the given graph as well as in the reverse graph.
A query can then be reduced to the set intersection problem.
Pruned-search-based approaches precompute information to speed up queries by
pruning~the~search.

Due to its volume, it is impossible to compare against all previous work.
We mostly follow the methodology of Merz and Sanders~\cite{merz2014preach} and focus
on five recent techniques.
The two most recent hop-labeling-based approaches are
\TF{}~\cite{DBLP:conf/sigmod/ChengHWF13} and
\PPL{}~\cite{DBLP:conf/cikm/YanoAIY13}.
In the pruned search category, the three most recent approaches are
\Preach{}~\cite{merz2014preach},
\IP{}~\cite{DBLP:journals/vldb/WeiYLJ18},
and \BFL{}~\cite{DBLP:journals/tkde/SuZWY17}.
We now go into more detail:

\textbf{\TF{}.}
The work by Cheng~\etal~\cite{DBLP:conf/sigmod/ChengHWF13} uses a data
structure called topological folding.
On the condensation \DAG{}, the authors define a topological structure that is
obtained by recursively folding the structure in half each time.
Using this topological structure, the authors create labels that help to
quickly answer reachability queries.

\textbf{\PPL{}.}
Yano~\etal~\cite{DBLP:conf/cikm/YanoAIY13} use pruned landmark labeling and
pruned path labeling as labels for their reachability queries.
In general, the
method follows the 2-hop labeling technique mentioned above, which stores sets
of vertices %
for each vertex $v$ and reduces queries to the set intersection problem.
Their techniques are able to reduce the size of the stored labels and hence to
improve query time and space consumption.

\textbf{\Preach{}.}
Merz and Sanders~\cite{merz2014preach} apply the approach of \emph{contraction
hierarchies}~(CHs)~\cite{geisberger2008contraction,geisberger2012exact} known from
shortest-path queries to the reachability problem.
The method first tries to answer queries by using pruning and precomputed
information such as topological levels (\Observation{basic:ts-level-fw}
and~\Ob{basic:ts-level-bw}).
It adopts and improves techniques from
\Grail{}~\cite{yildrim2010grail,yildirim2012grail} for that task, which
is distinctly outperformed by \Preach{} in the subsequent experiments.
Should these techniques not answer the query, \Preach{} instead performs a
bidirectional breadth-first search (BFS) using the computed hierarchy, \ie, for
a $\Query(s,t)$ the BFS only considers neighboring vertices with larger
topological level and along the CH.
The overall approach is simple and guarantees linear space and near linear
preprocessing time.

\textbf{\IP{}.}
Wei~\etal~\cite{DBLP:journals/vldb/WeiYLJ18} use a randomized labeling approach
by applying independent permutations on the labels.
Contrary to other labeling approaches, \IP{} checks for set-containment instead
of set-intersection.
Therefore, \IP{} tries to answer negative queries by checking for at least one
vertex that it is contained in only one of the two sets, where each set can
consist of at most \kIP{} vertices.
If this test fails, \IP{} checks another label, which contains precomputed
reachability information from the \hIP{} vertices with largest out-degree, and
otherwise
falls back to depth-first-search (DFS).

\textbf{\BFL{}.}
Su~\etal~\cite{DBLP:journals/tkde/SuZWY17} propose a labeling method which is
based on \IP{}, but additionally uses Bloom filters for storing and comparing
labels, which are then used to answer negative queries.
As parameters, \BFL{} accepts \sBFL{} and \dBFL{}, where $\sBFL{}$ denotes the length of
the Bloom filters stored for each vertex and $\dBFL{}$
controls the false positive rate.
By default, $\dBFL{} = 10\cdot \sBFL{}$.

\Table{algorithm-complexities}
subsumes the time and space complexities of
the new algorithm \Sreach{} that we introduce in \Section{algorithms} as well
as all algorithms mentioned in this paper except for \TF{}, where the
expressions describing the theoretical complexities are bulky and quite
complex themselves.

\section{O'Reach: Faster Reachability via Observations}%
\SectLabel{algorithms}
In this section we propose our new algorithm \Sreach{}, which is based on a set
of simple, yet powerful observations that enable us to answer a large
proportion of reachability queries in constant time and brings together
techniques from both hop labeling and pruned search.
Unlike regular hop-labeling-approaches, however, its initialization time is
linear.
As a further plus, our algorithm is configurable via multiple parameters and
extremely space-efficient with an index of only
$\SI[parse-numbers = false]{38n}{\Byte}$
in the most space-saving configuration that could
handle all instances used in \Section{experiments}
and uses all features.

\textbf{Overview.} %
The hop labeling technique used in our algorithm is inspired by a recent result
for experimentally faster reachability queries in a dynamic graph by
Hanauer~\ea{}~\cite{hanauer_et_al:LIPIcs:2020:12088}.
The idea here is to speed up reachability queries based on a selected set of
so-called \emph{supportive vertices}, for which complete out- and
in-reachability is maintained explicitly.
This information is used in three simple observations, which allow
to answer matching queries in constant time.
In our algorithm, we transfer this idea to the static setting.
We further increase the ratio of queries answerable in constant time by a new
perspective on topological orderings and their conflation with depth-first
search, which provides additional reachability information and further
increases the ratio of queries answerable in constant time.
In case that we cannot answer a query via an observation, we fall back to
either a pruning bidirectional breadth-first search or one of the existing
algorithms.

In the following, we switch the order and first discuss topological orderings
in depth, followed by our adaptation of supportive vertices.
For both parts, consider a reachability \Query($s, t$) for two vertices
$s, t \in V$ with $s \neq t$.

\subsection{Extended Topological Orderings}\SectLabel{extended-topsort}
Taking up on the observation that topological orderings can be used to answer a
reachability query decisively negative, we first investigate how
\Observation{basic:topsort} can be used most effectively in practice.
Before we dive deeper into this subject, let us briefly review some facts
concerning topological orderings and reachability in general.
\newbool{tsFactsProof}
\booltrue{tsFactsProof}%
\ifbool{tsFactsProof}{}{%
The proof of the following statement is simple and only given in the full
version~\cite{arxiv} for completeness.
}
\begin{theorem}\ThmLabel{ts-facts}
Let $\AnswerableNegQueries(\TopSort) \subseteq \NegQueries$ denote the set of
negative queries a topological ordering $\TopSort$ can answer, \ie, the set of
all $(s, t) \in \NegQueries$ such that $\TopSort(t) < \TopSort(s)$, and let
$\AnswerableNegQueryRatio(\TopSort) = \AnswerableNegQueries(\TopSort) /
\NegQueries$ be the answerable negative query ratio.
\begin{enumerate}[label=(\roman{*})]
\item\ItemLabel{ts-facts-unique}
The reachability in any \DAG{} is at most 50\%.
In this case, the topological ordering is unique.
\item\ItemLabel{ts-facts-50}
Any topological ordering $\TopSort$ witnesses the non-reachability between
exactly 50\% of all pairs of distinct vertices.
Therefore, $\AnswerableNegQueryRatio(\TopSort) \geq 50\%$.
\item\ItemLabel{ts-facts-equal}
Every topological ordering of the same \DAG{} can answer the same
\underline{ratio} of all negative queries via \Observation{basic:topsort}, \ie,
for two topological orderings $\TopSort$, $\TopSort'$:
$\AnswerableNegQueryRatio(\TopSort) = \AnswerableNegQueryRatio(\TopSort')$.
\item\ItemLabel{ts-facts-different}
For two different topological orderings $\TopSort \neq \TopSort'$ of a \DAG{},
$\AnswerableNegQueries(\TopSort) \neq \AnswerableNegQueries(\TopSort')$.
\end{enumerate}
\end{theorem}
\ifbool{tsFactsProof}{%
\begin{proof}
Let $G$ be a directed acyclic graph (DAG).
\begin{enumerate}%
\item[\Item{ts-facts-unique}]
As $G$ is acyclic, there is at least one topological ordering $\TopSort$ of
$G$.
Then, for every edge $(u, v)$ of $G$, $\TopSort(u) < \TopSort(v)$, which
implies that each vertex $u$ can reach at most all those vertices $w \neq u$
with $\TopSort(u) < \TopSort(w)$.
Consequently, a vertex $u$ with $\TopSort(u) = i$ can reach at most $n-i-1$
\emph{other} vertices (note that $i \geq 0$).
Thus, the reachability in $G$ is at most
$\frac{1}{n(n-1)}\sum_{i=0}^{n-1} (n-i-1)
= \frac{1}{n(n-1)}\sum_{j=0}^{n-1} j 
= \frac{n(n-1)}{n(n-1)\cdot 2} 
= \frac{1}{2}$.
Conversely, assume that the reachability in $G$ is $\frac{1}{2}$.
Then, each vertex $u$ with $\TopSort(u) = i$ reaches exactly all $n-i-1$ other
vertices ordered after it, which implies that there exists no other topological
ordering $\TopSort'$ with $\TopSort'(u) > \TopSort(u)$.
By induction on $i$, the topological ordering of $G$ is unique.
\item[\Item{ts-facts-50}]
Let $\TopSort$ be an arbitrary topological ordering of $G$.
Then, each vertex $u$ with $\TopSort(u) = i$ can certainly \emph{reach} those
vertices $v$ with $\TopSort(v) < \TopSort(u)$.
Hence, $\TopSort$ witnesses the non-reachability of exactly
$\sum_{i=1}^{n-1} i = \frac{n(n-1)}{2}$
pairs of distinct vertices.
\item[\Item{ts-facts-equal}]
As \Observation{basic:topsort} corresponds exactly to the non-reachability
between those pairs of vertices witnessed by the topological ordering, the claim
follows directly from \Item{ts-facts-50}.
\item[\Item{ts-facts-different}]
As $\TopSort \neq \TopSort'$, there is at least one $i \in \mathbb{N}_0$ such
that $\TopSort(u) = i = \TopSort'(v)$ and $u \neq v$.
Let $j = \TopSort(v)$.
If $j > i$, the number of non-reachabilities from $v$ to another vertex
witnessed by $\TopSort$ exceeds the number of those witnessed by $\TopSort'$,
and falls behind it otherwise.
In both cases, the difference in numbers immediately implies a difference in
the set of vertex pairs, which proves the claim.
\qedhere
\end{enumerate}
\end{proof}
}{}%
In consequence, it is pointless to look for one particularly good topological
ordering.
Instead, to get the most out of \Observation{basic:topsort}, we need
topological orderings whose sets of answerable negative queries differ greatly,
such that their union covers a large fraction of $\NegQueries$.
Note that both forward and backward topological levels each represent the set
of topological orderings that can be obtained by ordering the vertices in
blocks grouped by their level and arbitrarily permuting the vertices in each
block.
Different algorithms~\cite{Kahn62,Tarjan72,CLRS09} for computing a topological
ordering in linear time have been proposed over the years, with Kahn's
algorithm~\cite{Kahn62} in combination with a queue being one that always
yields a topological ordering represented by forward topological levels.
We therefore complement the forward and backward topological levels by
stack-based approaches, as in Kahn's algorithm~\cite{Kahn62} in combination
with a stack or Tarjan's DFS-based algorithm~\cite{Tarjan72} for computing the
\SCC{}s of a graph, which as a by-product also yields a topological ordering of
the condensation.
To diversify the set of answerable negative queries further, we additionally
randomize the order in which vertices are processed in case of ties and also
compute topological orderings on the reverse graph, in analogy to backward
topological levels.

We next show how, with a small extension, the stack-based topological orderings
mentioned above can be used to additionally answer positive queries.
To keep the description concise, we concentrate on Tarjan's
algorithm~\cite{Tarjan72} in the following and reduce it to the part
relevant for obtaining a topological ordering of a \DAG{}.
In short, the algorithm starts a depth-first search at an arbitrary vertex
$s \in S$, where $S \subseteq V$ is a given set of vertices to start from.
Whenever it visits a vertex $v$, it marks $v$ as visited and
recursively visits all unvisited vertices in its out-neighborhood.
On return, it \emph{prepends} $v$ to the topological ordering.
A loop over $S = V$ ensures that all vertices are visited.
Note that although the vertices are visited in DFS order, the topological
ordering is different from a DFS numbering as it is constructed ``from back to
front'' and corresponds to a reverse sorting according to what is also called
\emph{finishing time} of each vertex.

To answer positive queries, we exploit the invariant that when visiting a
vertex $v$, all yet unvisited vertices reachable from $v$ will be prepended to
the topological ordering prior to $v$ being prepended.
Consequently, $v$ can \emph{certainly} reach all vertices in the topological
ordering between $v$ and, exclusively, the vertex $w$ that was at the front of
the topological ordering when $v$ was visited.
Let $x$ denote the vertex preceding $w$ in the final topological ordering, \ie,
the vertex with the largest index that was reached recursively from $v$.
For a topological ordering $\TopSort$ constructed in this way, we call
$\TopSort(x)$ the \emph{high index} of $v$ and denote it with
$\High{\TopSort}(v)$.
Furthermore, $v$ \emph{may} be able to also reach $w$ and vertices beyond,
which occurs if $v \PathTo y$ for some vertex $y$, but $y$ had already been
visited earlier.
We therefore additionally track the \emph{max index}, the largest index of any
vertex that $v$ can reach, and denote it with $\Max{\TopSort}(v)$.
\Algorithm{topsortplus} shows how to compute an extended topological ordering
with both high and max indices in pseudo code and highlights our extensions.
Compared to Tarjan's original version~\cite{Tarjan72}, the running time remains
unaffected by our modifications and is still in $\bigO(n+m)$.
\begin{figure}[tb]
\centering
\begin{tikzpicture}
\node[inner sep=2pt,rectangle,draw] (alg) {%
\phantomsubcaption\label{alg:topsortplus}%
\includegraphics[width=.5\textwidth]{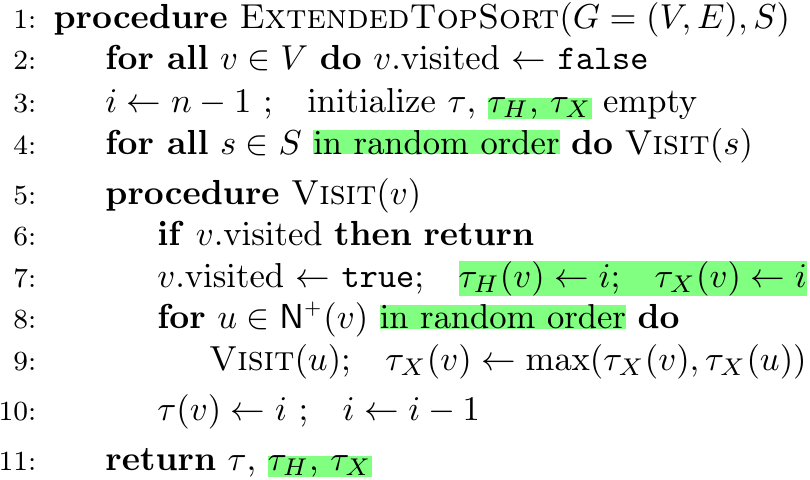}};
\node[inner sep=0pt,anchor=north west] (graphics) at ($(alg.north east) + (1cm,0)$) {
\phantomsubcaption\label{fig:extended-topsort}%
\includegraphics[width=.34\textwidth]{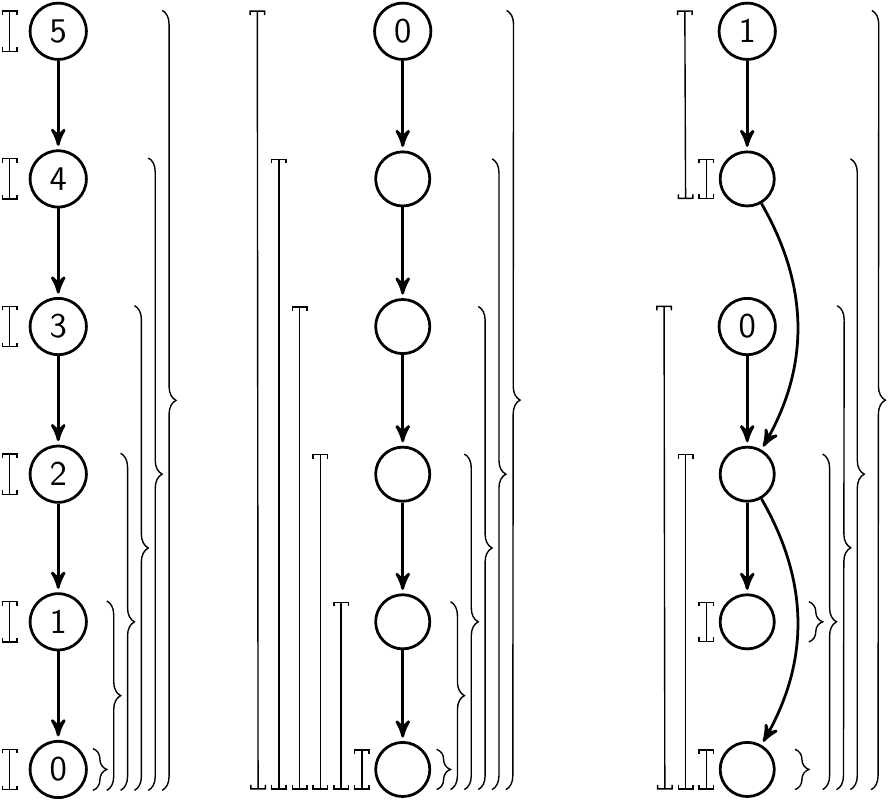}};
\node[anchor=north east,inner ysep=0pt,inner xsep=2pt,text height=2ex] at (alg.north west) {(\subref*{alg:topsortplus})}; 
\node[anchor=north east,inner ysep=0pt,inner xsep=2pt,text height=2ex] at (graphics.north west) {(\subref*{fig:extended-topsort})}; 
\end{tikzpicture}
\caption{%
(\subref{alg:topsortplus}):
Extended Topological Sorting.
(\subref{fig:extended-topsort}):
Three extended topological orderings of two graphs:
The labels correspond to the order in the start set $S$.
If the label is empty, the vertex need not be in $S$ or can have any larger
number.
The brackets to the left show the range $[\TopSort(v), \High{\TopSort}(v)]$,
the braces to the right the range $[\TopSort(v), \Max{\TopSort}(v)]$.
}%
\label{fig:extended-topsort-combined}%
\end{figure}

Note that neither max nor high indices yield an ordering of $V$:
Every vertex that is visited recursively starting from $v$ and before vertex
$x$ with $\TopSort(x) = \High{\TopSort}(v)$, inclusively, has the same high
index as $v$, and the high index of each vertex in a graph consisting of a
single path, \eg, would be $n-1$.
In particular, neither max nor high index form a DFS numbering and also differ
in definition and use from the DFS finishing times $\hat{\phi}$ used in
\Preach{}, where a vertex $v$ can \emph{certainly} reach vertices with DFS number up
to $\hat{\phi}$ and \emph{certainly none} beyond.
Conversely, $v$ may be able to also reach vertices with smaller DFS number than
its own, which cannot occur in a topological ordering.

If \textsc{ExtendedTopSort}
is run on the reverse graph, it yields a topological
ordering $\TopSort'$ and high and max indices $\High{\TopSort'}$ and
$\Max{\TopSort'}$, such that reversing $\TopSort'$ yields again a topological
ordering $\TopSort$ of the original graph.
Furthermore, $\Low{\TopSort}(v) := n - 1 - \High{\TopSort'}(v)$ is a
\emph{low index} for each vertex $v$, which denotes the smallest index of a
vertex in $\TopSort$ that can certainly reach $v$, \ie, the
out-reachability of $v$ is replaced by in-reachability.
Analogously, $\Min{\TopSort}(v) := n - 1 - \Max{\TopSort'}(v)$ is a \emph{min
index} in $\TopSort$ and no vertex $u$ with $\TopSort(u) <
\Min{\TopSort}(v)$ can reach $v$.

The following observations show how such an extended topological ordering
$\TopSort$ can be used to answer both positive and negative reachability
queries:
\medskip
\begin{enumerate}[label=(T\arabic{*})]
\addtolength{\itemindent}{1cm}
\begin{minipage}{.5\linewidth}
\item\ObLabel{tsp:hi}
If $\TopSort(s) \leq \TopSort(t) \leq \High{\TopSort}(s)$,
then $s \PathTo t$.
\item\ObLabel{tsp:max}
If %
$\TopSort(t) > \Max{\TopSort}(s)$,
then $s \NoPathTo t$.
\item\ObLabel{tsp:eqmax}
If %
$\TopSort(t) = \Max{\TopSort}(s)$,
then $s \PathTo t$.
\end{minipage}
\begin{minipage}{.5\linewidth}
\item\ObLabel{tsp:lo}
If $\Low{\TopSort}(t) \leq \TopSort(s) \leq \TopSort(t)$,
then $s \PathTo t$.
\item\ObLabel{tsp:min}
If %
$\TopSort(s) < \Min{\TopSort}(t)$,
then $s \NoPathTo t$.
\item\ObLabel{tsp:eqmin}
If %
$\TopSort(s) = \Min{\TopSort}(t)$,
then $s \PathTo t$.
\end{minipage}
\end{enumerate}
\smallskip
Recall that by definition, $\TopSort(s) \leq \High{\TopSort}(s) \leq
\Max{\TopSort}(s)$ and $\Min{\TopSort}(t) \leq \Low{\TopSort}(t) \leq
\TopSort(t)$.
\Figure{extended-topsort} depicts three examples for extended topological
orderings.
In contrast to negative queries, not every extended topological ordering is
equally effective in answering positive queries, and it can be arbitrarily bad,
as shown in the extremes on the left (worst) and at the center (best) of
\Figure{extended-topsort}:
\begin{theorem}
Let $\AnswerablePosQueries(\TopSort) \subseteq \PosQueries$ be the set of
positive queries an extended topological ordering $\TopSort$ can answer and let
$\AnswerablePosQueryRatio(\TopSort) = \AnswerablePosQueries(\TopSort) /
\PosQueries$ be the answerable positive query ratio.
Then, $0 \leq \AnswerablePosQueryRatio(\TopSort) \leq 1$.
\end{theorem}
Instead, the effectiveness of an extended topological ordering depends
positively on the
size of the ranges $\Range{\TopSort(v), \High{\TopSort}(v)}$ and
$\Range{\Low{\TopSort}(v), \TopSort(v)}$,
and negatively on $\Range{\High{\TopSort}(v), \Max{\TopSort}(v)}$
and $\Range{\Min{\TopSort}(v), \Low{\TopSort}(v)}$
which in turn depend on the recursion depths during construction and the order
of recursive calls.
The former two can be maximized if the first, non-recursive call to
\Call{Visit}{} in line~4
in %
\textsc{ExtendedTopSort}
always has a source as its argument, \ie, if the algorithm's parameter $S$
corresponds to the set of all sources.
Clearly, this still guarantees that every vertex is visited.

In addition to the forward and backward topological levels, \Sreach{} thus
computes a set of $\ParamNumTopsorts$ extended topological orderings starting
from sources, where $\ParamNumTopsorts$ is a tuning parameter, and
$\ParamNumTopsorts/2$ of them are obtained via the reverse graph.
It then applies \Observation{basic:topsort} as well as
Observations~\Ob{tsp:hi}--\Ob{tsp:eqmin} to all extended topological orderings.
\subsection{Supportive Vertices}\SectLabel{supportive-vertices}
We now show how to apply and improve the idea of supportive vertices in the
static setting.
A vertex $v$ is \emph{supportive} if the set of vertices that $v$ can reach and
that can reach $v$, $\OutReachSet(v)$ and $\InReachSet(v)$, respectively, have
been precomputed and membership queries can be performed in sublinear time.
We can then answer reachability queries using the following simple
observations~\cite{hanauer_et_al:LIPIcs:2020:12088}:
\begin{enumerate}[label=(S\arabic{*})]
\addtolength{\itemindent}{1cm}
\item\ObLabel{sv:in-s-out-t}
If $s\in\InReachSet(v)$ and $t\in\OutReachSet(v)$ for \emph{any} $v \in V$, then $s \PathTo t$.
\item\ObLabel{sv:out-s-nout-t}
If $s\in\OutReachSet(v)$ and $t\not\in\OutReachSet(v)$ for \emph{any} $v \in V$, then $s \NoPathTo t$.
\item\ObLabel{sv:nin-s-in-t}
If $s\not\in\InReachSet(v)$ and
$t\in\InReachSet(v)$ for \emph{any} $v \in V$, then $s \NoPathTo t$.
\end{enumerate}
To apply these observations, our algorithm selects a set of $\ParamNumSupports$
supportive vertices during the initialization phase.
In contrast to the original use scenario in the dynamic setting, where the
graph changes over time and it is difficult to choose ``good'' supportive
vertices that can help to answer many queries, the static setting leaves room
for further optimizations here:
With respect to \Observation{sv:in-s-out-t}, we consider a supportive vertex
$v$ ``good'' if $|\OutReachSet(v)| \cdot |\InReachSet(v)|$ is large as
it maximizes the possibility that $s \in \InReachSet(v) \wedge t \in
\OutReachSet(v)$.
With respect to \Observation{sv:out-s-nout-t} and~\Ob{sv:nin-s-in-t}, we expect
a ``good'' supportive vertex to have out- or in-reachability sets,
respectively, of size close to $\frac{n}{2}$, i.e., when
$|\OutReachSet(v)|\cdot|V \setminus \OutReachSet(v)|$ or
$|\InReachSet(v)|\cdot|V \setminus \InReachSet(v)|$, respectively, are maximal.
Furthermore, to increase total coverage and avoid redundancy, the set of
queries \Query($s, t$) covered by two different supportive vertices should
ideally overlap as little as possible.

\Sreach{} takes a parameter $\ParamNumSupports$ specifying the number of
supportive vertices to pick.
Intuitively speaking, we expect vertices in the topological ``mid-levels'' to
be better candidates than those at the ends, as their out- and
in-reachabilities (or non-reachabilities) are likely to be more balanced.
Furthermore, if \emph{all} vertices on one forward (backward) level
$i$ were supportive, then \emph{every} \Query($s, t$) with
$\TopSortLevelForward(s) < i < \TopSortLevelForward(t)$
($\TopSortLevelBackward(t) < i < \TopSortLevelBackward(s)$) could be answered
using only \Observation{sv:in-s-out-t}.
As finding a ``perfect'' set of supportive vertices is computationally
expensive and we strive for linear preprocessing time, we experimentally
evaluated different strategies for the selection process.
Due to page limits, we only describe the most successful one:
A forward (backward) level $i$ is called \emph{central}, if
$\frac{1}{5}\TopSortLevel_{\max} \leq i \leq \frac{4}{5}\TopSortLevel_{\max}$,
where $\TopSortLevel_{\max}$ is the maximum topological level.
A level $i$ is called \emph{slim} if there at most $\ParamLevelThreshold$
vertices having this level, where $\ParamLevelThreshold$ is a parameter to
\Sreach{}.
We first compute a set of candidates of size at most $\ParamNumSupports \cdot
\ParamSupportSampleSize$ that contains all vertices on
slim forward or backward levels, arbitrarily discarding vertices as soon as the
threshold $\ParamNumSupports \cdot \ParamSupportSampleSize$ is reached.
$\ParamSupportSampleSize$ is another parameter to \Sreach{} and together with
$\ParamNumSupports$ controls the size of the candidate set.
If the threshold is not reached, we fill up the set of candidates by picking
the missing number of vertices uniformly at random from all other vertices
whose forward level is central.
In the next step, the out- and in-reachabilities of all candidates are obtained
and the $\ParamNumSupports$ vertices $v$ with largest $|\OutReachSet(v)| \cdot
|\InReachSet(v)|$ are chosen as supportive vertices.
This strategy primarily optimizes for \Observation{sv:in-s-out-t}, but worked
better in experiments than strategies that additionally tried to optimize for
\Observation{sv:out-s-nout-t} and~\Ob{sv:nin-s-in-t}.
The time complexity of this process is in
$\bigO(\ParamNumSupports\ParamSupportSampleSize(n+m) +
\ParamNumSupports\ParamSupportSampleSize\log(\ParamNumSupports\ParamSupportSampleSize))$.

We remark that this is a general-purpose approach that has shown to work well
across different types of instance, albeit possibly at the expense of an
increased initialization time.
It seems natural that more specialized routines for different graph classes can
improve both running time and coverage.
\subsection{The Complete Algorithm}
Given a graph $G$ and a sequence of queries $Q$,
we summarize in the following how \Sreach{} proceeds.
During initialization, it performs the following steps:
\begin{enumerate}[label={Step \arabic{*}:},ref={\arabic{*}},labelindent=.5\parindent,leftmargin=*]
\item\ItemLabel{init-cc}
Compute the \CC{}s
\item\ItemLabel{init-toplevels}
Compute forward/backward topological levels
\item\ItemLabel{init-exttopord}
Obtain $\ParamNumTopsorts$ random extended topological orderings
\item\ItemLabel{init-supports}
Pick $\ParamNumSupports$ supportive vertices, compute $\OutReachability{\cdot}$ and $\InReachability{\cdot}$
\end{enumerate}
Steps~\Item{init-cc} and~\Item{init-toplevels} run in linear time.
As shown in \Section{extended-topsort}
and \Section{supportive-vertices}, the same applies to
Steps~\Item{init-exttopord} and~\Item{init-supports}, assuming that all
parameters are constants.
The required space is linear for all steps.
The reachability index consists of the following information for each vertex $v$:
one integer for the \CC{},
one integer each for
$\TopSortLevelForward(v)$ and $\TopSortLevelBackward(v)$,
three integers for each of the $\ParamNumTopsorts$ extended topological
orderings $\TopSort$ ($\TopSort(v), \High{\TopSort}(v)/\Low{\TopSort}(v),
\Max{\TopSort}(v)/\Min{\TopSort}(v)$),
two bits for each of the $\ParamNumSupports$ supportive vertices,
indicating its reachability to/from $v$.
For graphs with $n \leq 2^{32}$, \SI{4}{\Byte} per integer suffice.
Furthermore, we group the bits encoding the reachabilities to and from the
supportive vertices, respectively, and represent them each by one suitably sized
integer, \eg, using \texttt{uint8\_t} (\SI{8}{\bit}), for $\ParamNumSupports
\leq 8$ supportive vertices.
As the smallest integer has at least \SI{8}{\bit} on most
architectures, we store \SI[parse-numbers = false]{12 + 12 \ParamNumTopsorts{} +
2 \cdot \lceil\frac{\ParamNumSupports}{8}\rceil}{\Byte} per vertex.

For each query \Query($s, t$), \Sreach{} tries to answer it using one of the
observations in the order given below, which on the one hand has been optimized
by some preliminary experiments on a small subset of benchmark instances (see
\Section{experiments} for details) and on the other hand strives for a fair
alternation between ``positive'' and ``negative'' observations to avoid
overfitting.
Note that all observation-based tests run in constant time.
As soon as one of them can answer the query affirmatively, the
result is returned immediately.
A test leading to a positive or negative answer is marked as \FastTrueAnswer{}
or \FastFalseAnswer{}, respectively.
\begin{enumerate}[label=Test \arabic{*}:,ref={\arabic{*}},labelindent=.5\parindent,leftmargin=*]
\item\ItemLabel{test:s-eq-t} \FastTrueAnswer{}
$s = t$?
\item\ItemLabel{test:toplevels}
\FastFalseAnswer{}
\FastFalseAnswer{}
topological levels
\Ob{basic:ts-level-fw}, \Ob{basic:ts-level-bw}
\item\ItemLabel{test:sv-pos} \FastTrueAnswer{}
$\ParamNumSupports$ supportive vertices, positive
\Ob{sv:in-s-out-t}
\item\ItemLabel{test:ts-first}
\FastFalseAnswer{}
\FastTrueAnswer{}
\FastFalseAnswer{}
\FastTrueAnswer{}
first topological ordering
\Ob{basic:topsort}, \Ob{tsp:hi}, \Ob{tsp:max}, \Ob{tsp:eqmax}
\item\ItemLabel{test:sv-neg}
\FastFalseAnswer{}
\FastFalseAnswer{}
$\ParamNumSupports$ supportive vertices, negative
\Ob{sv:out-s-nout-t}, \Ob{sv:nin-s-in-t}
\item\ItemLabel{test:ts-rem}
\FastFalseAnswer{}
\FastTrueAnswer{}
\FastFalseAnswer{}
\FastTrueAnswer{}
remaining $\ParamNumTopsorts-1$ topological orderings
\Ob{basic:topsort},
\Ob{tsp:hi}/\Ob{tsp:lo},
\Ob{tsp:max}/\Ob{tsp:min},
\Ob{tsp:eqmax}/\Ob{tsp:eqmin}
\item\ItemLabel{test:cc} \FastFalseAnswer{}
different \CC{}s
\Ob{basic:cc}
\end{enumerate}
Observe that the tests for \Observation{sv:in-s-out-t},~\Ob{sv:out-s-nout-t},
and~\Ob{sv:nin-s-in-t} can each be implemented easily using boolean logic,
which allows for a concurrent test of all supports whose reachability
information is encoded in one accordingly-sized integer:
For \Observation{sv:in-s-out-t}, it suffices to test whether
$r^-(s) \wedge r^+(t) > 0$,
and $r^+(s) \wedge \neg r^+(t) > 0$
and $\neg r^-(s) \wedge r^-(t) > 0$ for
Observations~\Ob{sv:out-s-nout-t} and~\Ob{sv:nin-s-in-t}, where $r^+$
and $r^-$ hold the respective forward and backward reachability
information in the same order for all supports.
Each test hence requires at most one comparison of two integers plus at most
two elementary bit operations.
Also note that Observation~\Ob{basic:sink} is
implicitly tested by Observations~\Ob{basic:ts-level-fw}
and~\Ob{basic:ts-level-bw}.
Using the data structure described above, our algorithm requires at most one
memory transfer for $s$ and one for $t$ for each \Query($s, t$) that is
answerable by one of the observations.
Note that there are more observations that allow to identify a negative query
than a positive query, which is why we expect a more pronounced speedup for the
former.
However, as stated in \Theorem{ts-facts}, the reachability in DAGs is
always less than \SI{50}{\percent}, which justifies a bias towards
an optimization for negative queries.

If the query can not be answered using any of these tests, we instead fall back
to either another algorithm or a bidirectional BFS with
pruning, which uses these tests for each newly encountered vertex $v$ in a
subquery \Query{}($v, t$) (forward step) or \Query{}($s, v$) (backward step).
If a subquery can be answered decisively positive by a test, the
bidirectional BFS can immediately answer \Query{}($s, t$) positively.
Otherwise, if a subquery is answered decisively negative by a test,
the encountered vertex $v$ is no longer considered (pruning step).
If the subquery could not be answered by a test, the vertex $v$ is
added to the queue as in a regular (bidirectional) BFS.

\section{Experimental Evaluation}\SectLabel{experiments}
We evaluated our new algorithm \Sreach{} as a preprocessor to various recent state-of-the-art
algorithms listed below against running these algorithms on their own.
Furthermore, we use as an additional fallback solution the pruned bidirectional
BFS~(\pBFS{}).
Our experimental study
follows the methodology in~\cite{merz2014preach}
and comprises the algorithms
\PPL{}~\cite{DBLP:conf/cikm/YanoAIY13},
\TF{}~\cite{DBLP:conf/sigmod/ChengHWF13},
\Preach{}~\cite{merz2014preach},
\IP{}~\cite{DBLP:journals/vldb/WeiYLJ18}, and
\BFL{}~\cite{DBLP:journals/tkde/SuZWY17}.
Moreover, our evaluation is the first that directly relates \IP{} and \BFL{} to
\Preach{} and studies the performance of \IP{} and \BFL{} separately for
successful (\Instance{positive}) and unsuccessful (\Instance{negative})
reachability queries.
For reasons of comparison, we also assess the query performance of a full
reachability matrix by computing the transitive closure of the input graph
entirely during initialization, storing it in a matrix using \SI{1}{\bit}
per pair of vertices, and answering each query by a single memory lookup.
We refer to this algorithm simply as \FullMatrix{}.
As the reachability in \DAG{}s is small and cache locality can influence lookup
times, we also experimented with various hash set implementations.
However, none was faster or more memory-efficient than \FullMatrix{}.
\ifbool{longversion}{%
\begin{table}[tbh]
\centering
\caption{Instances used in our experiments (read /\num{e3}: in thousands).
$\Sources$\%/$\Sinks$\%/$\Isolated$\%:
ratios of (non-isolated) sources/sinks, and isolated vertices.
\#\CC{}s(large): \#\CC{}s total(\#\CC{}s with at least $\frac{n}{10}$ vertices).
$\TopSortLevel_{\max}$: maximum topological forward/backward level,
equals the diameter.
$\ReachabilityRatio$: reachability by experiments.}%
\includegraphics[width=\linewidth,height=.93\textheight,keepaspectratio]{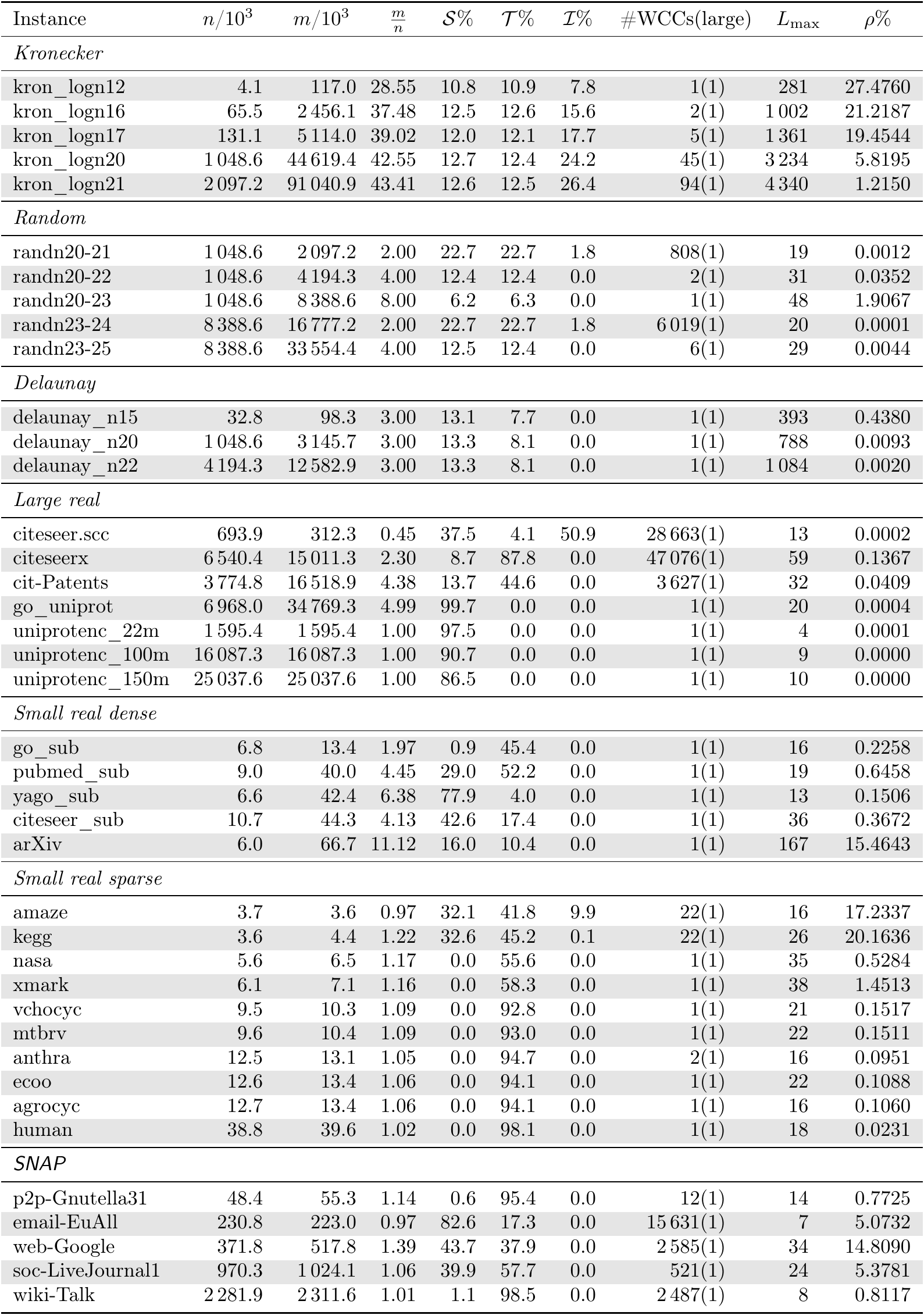}
\TabLabel{instances}
\end{table}
}{}

\textbf{Setup and Methodology.}
We implemented \Sreach{} in \Cpp{}14%
\footnote{We plan to release the code publicly.}
with \pBFS{} as built-in fallback strategy.
For
\PPL{}\footnote{Provided directly by the authors.}%
\addtocounter{footnote}{-1}\addtocounter{Hfootnote}{-1},
\TF{}\footnotemark,
\Preach{}%
\footnote{\url{https://github.com/fiji-flo/preach2014/tree/master/original_code}},
\IP{}%
\footnote{\url{https://github.com/datourat/IP-label-for-graph-reachability}}, and
\BFL{}%
\footnote{\url{https://github.com/BoleynSu/bfl}}
we used the original \Cpp{} implementation in each case.
All source code was compiled with \Tool{GCC} 7.5.0 and full optimization
(\texttt{-O3}).
The experiments were run on a Linux machine under Ubuntu 18.04 with kernel 4.15
on four AMD Opteron 6174 CPUs clocked at \SI{2.2}{\giga\hertz} with
\SI{512}{\kilo\byte} and \SI{6}{\mega\byte} L2 and L3 cache, respectively and \num{12} cores per CPU.
Overall, the machine has \num{48} cores and a total of \SI{256}{\giga\byte} of
RAM. %
Unless indicated otherwise, each experiment was run sequentially and
exclusively on one processor and its local memory. %
As non-local memory accesses incur a much higher cost, an exception to this
rule was only made for \FullMatrix{}, where we would otherwise have been able to
only run twelve instead of \num{29} instances.
We also parallelized the initialization phase for \FullMatrix{}, where the
transitive closure is computed, using \num{48} threads.
However, all queries were processed sequentially.

To counteract artifacts of measurement and accuracy, we ran each algorithm five
times on each instance and use the median for the evaluation.
As \Sreach{} uses randomization during initialization, we %
instead report the average running time over five different seeds.
For %
\IP{} and \BFL{}, which are randomized in the same way, but don't
accept a seed, we just give the average over five repetitions.
We note that also taking the median instead or increasing the number of
repetitions does not change the overall picture.

\textbf{Instances.}
To facilitate comparability, we adopt the instances used in the
papers introducing
\Preach{}~\cite{merz2014preach} %
and \TF{}~\cite{DBLP:conf/sigmod/ChengHWF13},
which overlap with those used to evaluate
\IP{}~\cite{DBLP:journals/vldb/WeiYLJ18} and
\BFL{}~\cite{DBLP:journals/tkde/SuZWY17},
and
which are available either from the \Grail{} code
repository%
\footnote{\url{https://code.google.com/archive/p/grail/}\FootLabel{grail}}
or the Stanford
Network Analysis Platform \SNAP{}~\cite{snapnets}.
Furthermore, we extended the set of benchmark graphs by further instance sizes
and Delaunay graphs.
\ifbool{longversion}{%
\Table{instances} provides a detailed overview.}{%
\Table{results-init-plus} provides a short overview,
more details are available in the full version~\cite{arxiv}.
}
As we only consider \DAG{}s, all instances are condensations of their
respective originals, if they were not acyclic already.
We also adopt the grouping of the instances as
in~\cite{yildirim2012grail,merz2014preach} and provide only a short description
of the different sets in the following.

\emph{Kronecker:}
These instances were generated by the RMAT generator for the
Graph500 benchmark~\cite{graph500}
and oriented acyclically from smaller to larger node ID.
The name encodes the number of vertices $2^i$ as \Instance{kron\_logn$i$}.
\emph{Random:}
Graphs generated according to the Erd\H{o}s-Renyí model $G(n, m)$ and
oriented acyclically from smaller to larger node ID.
The name encodes $n=2^i$ and $m=2^j$ as \Instance{randn$i$-$j$}.
\emph{Delaunay:}
Delaunay graphs from the 10th DIMACS
Challenge~\cite{benchmarksfornetworksanalysis,funke2017communication}.
\Instance{delaunay\_n$i$} is a Delaunay triangulation of $2^{i}$ random points
in the unit square.
\emph{Large real:}
Introduced in~\cite{yildirim2012grail}, these instances represent citation
networks (\Instance{citeseer.scc}, \Instance{citeseerx},
\Instance{cit-Patents}),
a taxonomy graph
(\Instance{go-uniprot}),
as well as excerpts from the RDF graph of a protein database
(\Instance{uniprotm22}, \Instance{uniprotm100}, \Instance{uniprotm150}).
\emph{Small real dense:}
Among these instances, introduced in~\cite{JXRF09-3hop},
are again citation networks (\Instance{arXiv}, \Instance{pubmed\_sub},
\Instance{citeseer\_sub}), a taxonomy graph (\Instance{go\_sub}),
as well as one obtained from a semantic knowledge database %
(\Instance{yago\_sub}).
\emph{Small real sparse:}
These instances were introduced in~\cite{DBLP:conf/sigmod/JinXRW08} and
represent XML documents (\Instance{xmark}, \Instance{nasa}), metabolic networks
(\Instance{amaze}, \Instance{kegg}) or originate from pathway and genome
databases (all others).
\emph{\SNAP{}:}
The e-mail network graph (\Instance{email-EuAll}),
peer-to-peer network (\Instance{p2p-Gnutella31}),
social network (\Instance{soc-LiveJournal1}),
web graph (\Instance{web-Google}),
as well as the communication network (\Instance{wiki-Talk})
are part of \SNAP{} and were first used in~\cite{DBLP:conf/sigmod/ChengHWF13}.

\textbf{Queries.}
Following the methodology of~\cite{merz2014preach}, we generated three sets of
\num{100000} queries each: \Instance{positive}, \Instance{negative}, and
\Instance{random}.  Each set consists of random queries, which were generated
by picking two vertices uniformly at random and filtering out negative or
positive queries for the \Instance{positive} and \Instance{negative} query
sets, respectively.
The fourth query set, \Instance{mixed}, is a randomly shuffled union of all
queries from \Instance{positive} and \Instance{negative} and hence contains
\num{200000} pairs of vertices.
As the order of the queries within each set had an observable effect on the
running time due to caching effects and memory layout, we randomly shuffled
every query set five times and used a different permutation for each repetition
of an experiment to ensure equal conditions for all algorithms.
\clearpage 

\subsection{Experimental Results}
We ran \Sreach{} with $\ParamNumSupports = 16$ supportive vertices, picked from
\num{1200} candidates ($\ParamSupportSampleSize = 75$,
$\ParamLevelThreshold=8$) and $\ParamNumTopsorts = 4$ extended topological
orderings.
We ran \IP{} with the two configurations used also by the
authors~\cite{DBLP:journals/vldb/WeiYLJ18} and refer to the resulting
algorithms as \IPsparse{} (\emph{sparse}, $\hIP{}= \kIP{} = 2$) and
\IPdense{} (\emph{dense}, $\hIP{}= \kIP{} = 5$).
Similarly, we evaluated \BFL{}~\cite{DBLP:journals/tkde/SuZWY17} with configuration
\emph{sparse} as \BFLsparse{} ($\sBFL{}=64$)
and \emph{dense} as \BFLdense{} ($\sBFL{}=160$),
following the presets given by the authors.

\begin{table}[tb]
\centering
\caption{%
Average query time per algorithm and query set.}%
\TabLabel{results-averages}
\includegraphics[width=\linewidth]{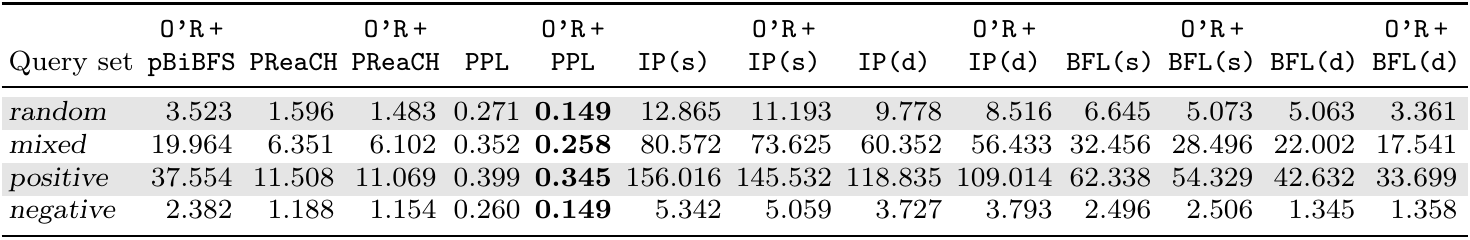}
\end{table}

\ifbool{longversion}{%
\afterpage{%
\begin{landscape}
\begin{table}
\centering
\caption{%
Average query times in \si{\micro\second} for \num{100000} negative (left) and positive queries (right).
Highlighted results are the overall best/second-best after \FullMatrix{} per query set over \emph{all} tested algorithms.}%
\TabLabel{results-negative-positive}
\includegraphics[width=.9\linewidth]{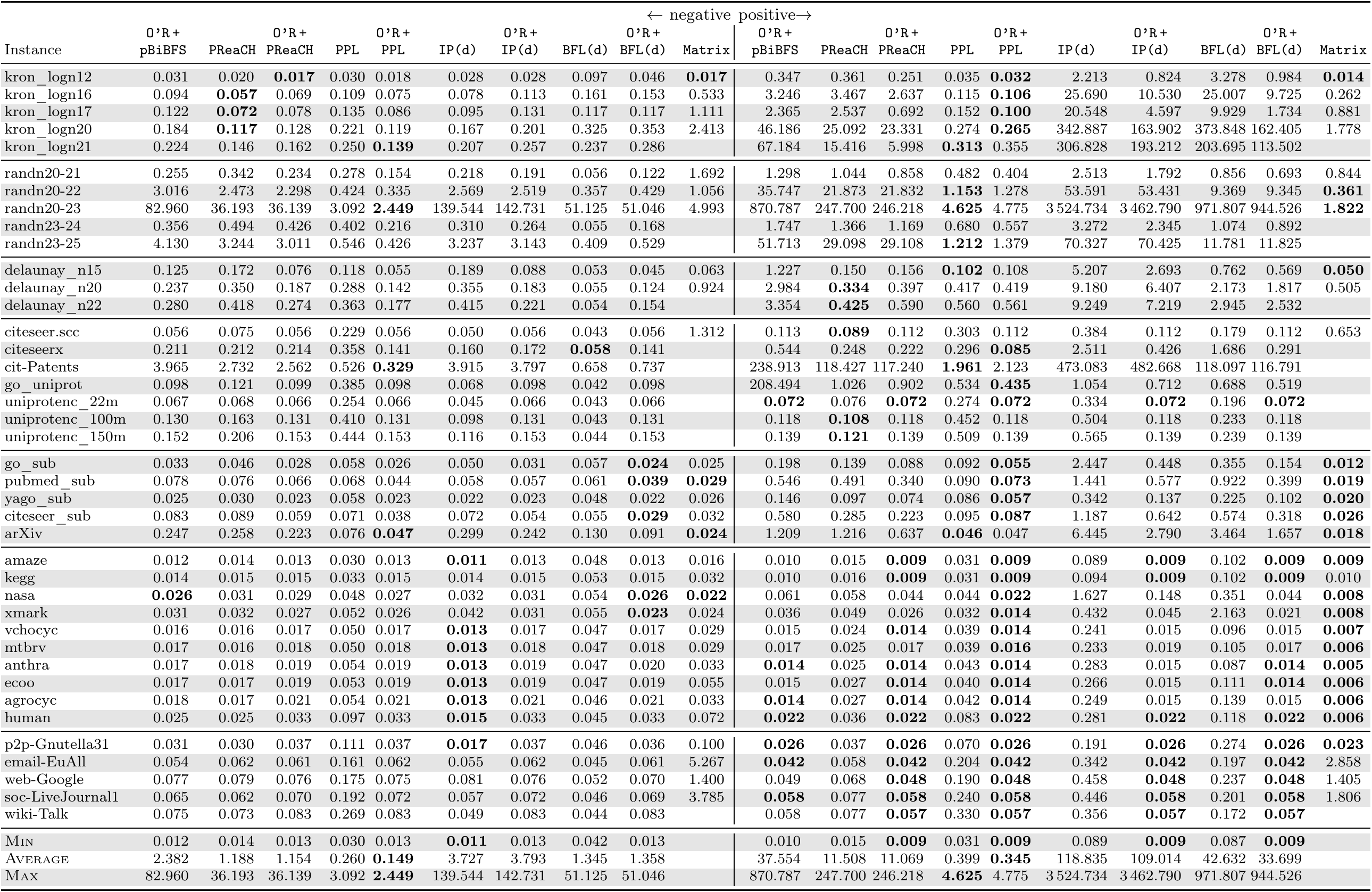}
\end{table}
\end{landscape}
}
\begin{table}[htb]
\centering
\caption{%
Average query times in \si{\micro\second} for \num{100000} negative (left) and positive queries (right).
Highlighted results are the overall best/second-best after \FullMatrix{} per query set over \emph{all} tested algorithms.}%
\TabLabel{results-negative-positive-rest}
\includegraphics[width=\linewidth]{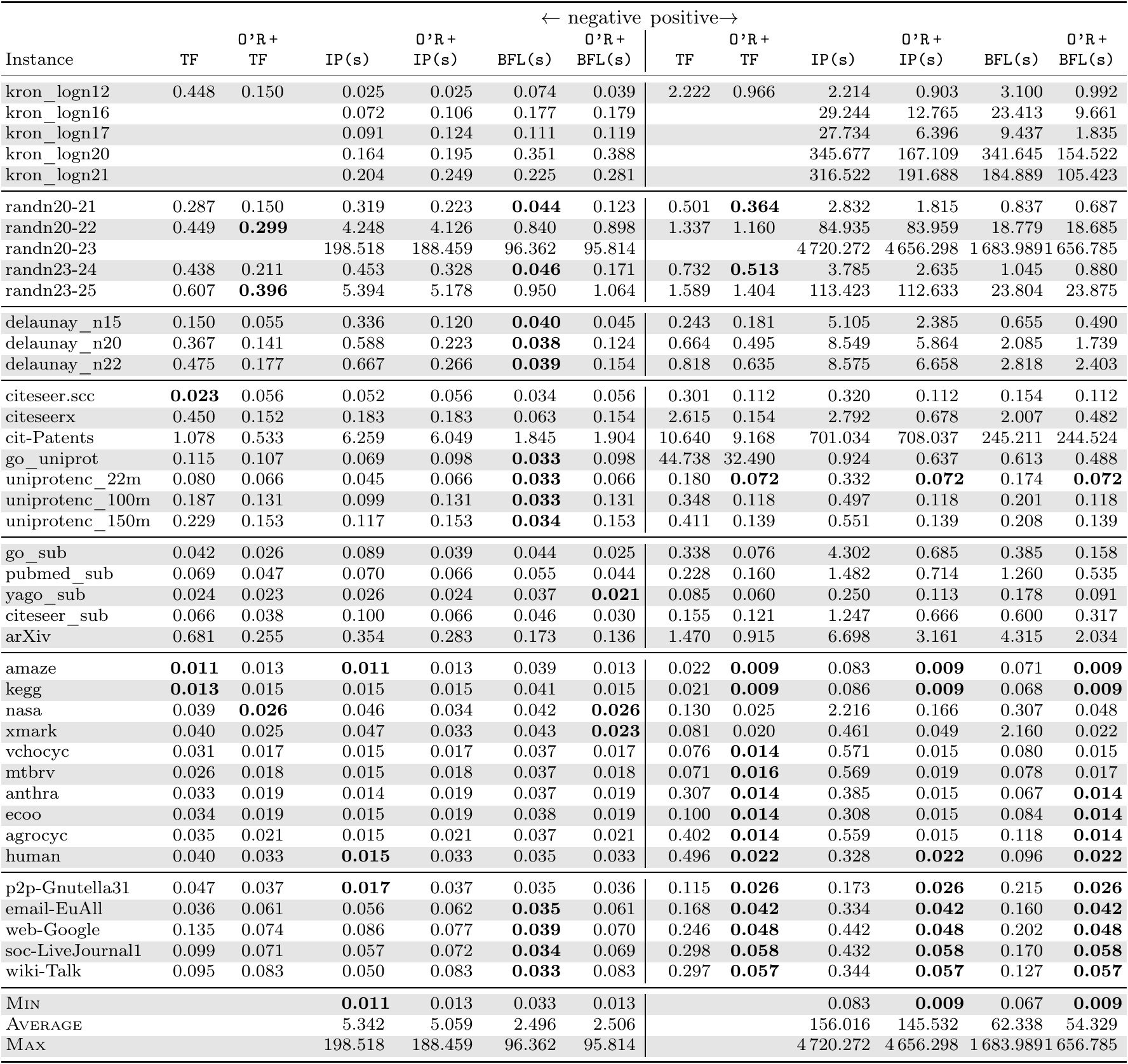}
\end{table}
}{}%
\textbf{Average query times.}
\Table{results-negative-positive} lists the average time per query for the
query sets \Instance{negative} and \Instance{positive}.
All missing values are due to a memory requirement of more than
\SI{32}{\giga\byte} (\TF{}) and \FullMatrix{} (\SI{256}{\giga\byte}).
For each instance and query set, the running time of the fastest algorithm is
printed in bold.
If \FullMatrix{} was fastest, also the running time of the second-best
algorithm is highlighted.
Besides \FullMatrix{}, the table shows
the running times of \Preach{}, \PPL{}, \IPdense{}, and \BFLdense{} alone
as well as multiple versions for \Sreach{}:
one with a pruned bidirectional BFS (\SrPlus{}\pBFS{}) as fallback as well as
one per competitor (\SrPlus{}\dots), where \Sreach{} was run without fallback
and the queries left unanswered were fed to the competitor.
\ifbool{longversion}{%
Analogously, the running times for
\IPdense{}, \BFLdense{}, and \TF{} alone and as fallback for \Sreach{}
are given in \Table{results-negative-positive-rest}.
}{%
Due to lack of space, detailed per-instance as well as average running times for \TF{},
\IPsparse{}, and \BFLsparse{} alone and as fallbacks are only given in the full
version~\cite{arxiv}.
}

Our results by and large \emph{confirm} the performance comparison of \Preach{}
\ifbool{longversion}{%
\PPL{}, and \TF{}
}{%
and \PPL{} (and \TF{}%
, see~\cite{arxiv})
}
conducted by Merz and Sanders~\cite{merz2014preach}.
\Preach{} was the fastest on three out of five Kronecker graphs for the
negative query set, once beaten by \SrPlus{}\Preach{} and \SrPlus{}\PPL{} each,
whereas \PPL{} and \SrPlus{}\PPL{} dominated all others on the positive query
set in this class as well as on three of the five random graphs, while
\SrPlus\TF{} was slightly faster on the other two. %
\ifbool{longversion}{%
In contrast to the study in~\cite{merz2014preach}, \TF{} is outperformed
slightly by \PPL{} on random instances for the positive query set.
}{}
\Preach{} was also the dominating approach on the small real sparse and
\SNAP{} instances in the aforementioned study~\cite{merz2014preach}.
By contrast, it was \emph{outperformed} on these classes here by \Sreach{} with
almost any fallback on all instances for the positive query set, and by either
\IPdense{} or \BFLsparse{} on almost all instances for the negative query set.
On the Delaunay and large real instances, \BFLsparse{} often was the fastest
algorithm on the set of negative queries.
The results also reveal that \BFL{} and in particular \IP{} have a weak spot in
answering positive queries.
\emph{On average over all instances}, \SrPlus{}\PPL{} had the \emph{fastest
average query time} both for \Instance{negative} and \Instance{positive}
queries.

Notably, $\FullMatrix{}$ was \emph{outperformed} quite often, especially for
queries in the set \Instance{negative}, which correlates with the fact that a
large portion of these queries could be answered by constant-time observations
(see also the detailed analysis of observation effectiveness below)
and is due to its larger memory footprint.
Across all instances and seeds, more than \SI{95}{\percent} of all queries
in this set could be answered by \Sreach{} directly.
On the set \Instance{positive}, the average query time for \FullMatrix{} was in
almost all cases less than on the \Instance{negative} query set, which is
explained by the small reachability of the instances and a resulting higher
spatial locality and better cacheability of the few and naturally clustered
one-entries in the matrix.
Consequently, this effect was distinctly reduced for the \Instance{mixed}
query set, as shown in \Table{results-random-mixed}.

There are some instances where \Sreach{} had a fallback rate of
over \SI{90}{\percent} for the \Instance{positive} query set,
\eg, on \Instance{cit-Patents}, which is clearly reflected in the running time.
Except for \PPL{}, all algorithms had difficulties with positive queries on
this instance.
Conversely, the fallback rate on all \Instance{uniprotenc\_${}^*$} instances
and \Instance{citeseer.scc}, e.g., was \SI{0}{\percent}.
On average across all instances and seeds, \Sreach{} could answer over
\SI{70}{\percent} of all \Instance{positive} queries by constant-time
observations.

\ifbool{longversion}{%
\afterpage{%
\begin{landscape}
\begin{table}
\centering
\caption{%
Average query times in \si{\micro\second} for \num{100000} random (left) and \num{200000} mixed queries (right).
Highlighted results are the overall best/second-best after \FullMatrix{} per query set over \emph{all} tested algorithms.}%
\TabLabel{results-random-mixed}
\includegraphics[width=.9\linewidth]{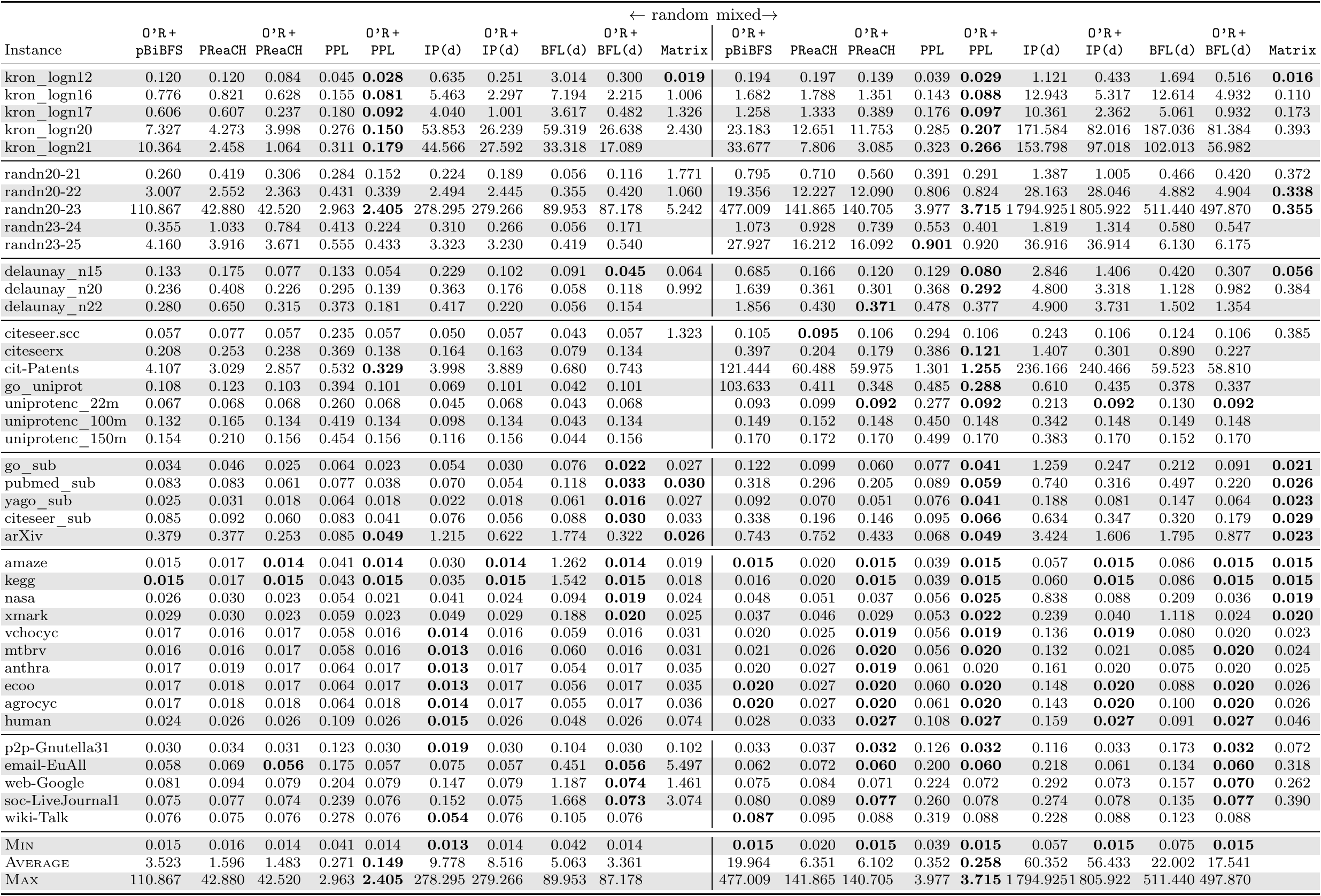}
\end{table}
\end{landscape}
}
\begin{table}
\centering
\caption{%
Average query times in \si{\micro\second} for \num{100000} random (left) and \num{200000} mixed queries (right).
Highlighted results are the overall best/second-best after \FullMatrix{} per query set over \emph{all} tested algorithms.}%
\TabLabel{results-random-mixed-rest}
\includegraphics[width=\linewidth]{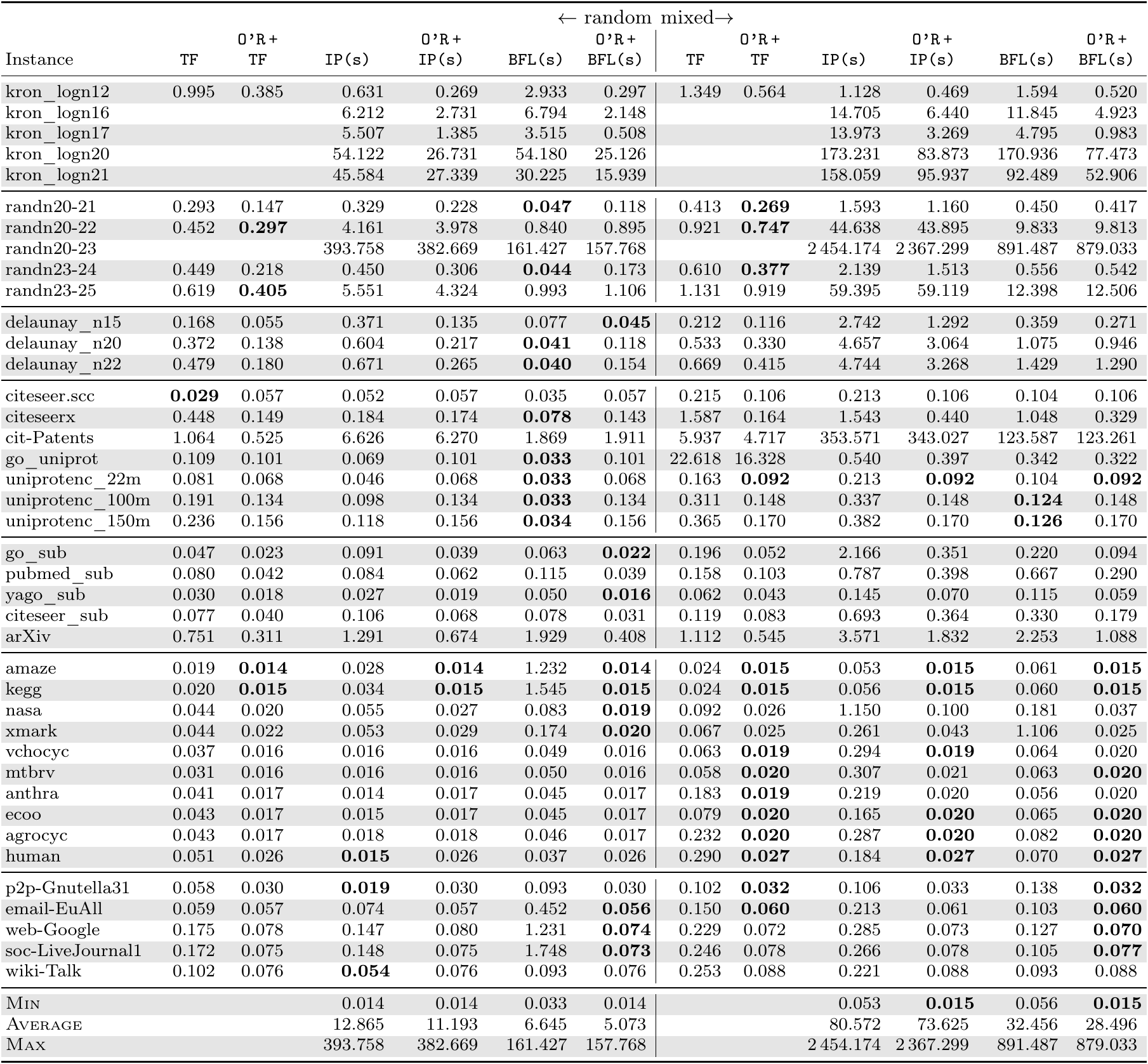}
\end{table}
}{}%
The results on the query sets \Instance{random} and \Instance{mixed} are
similar and listed in
\ifbool{longversion}{%
\Table{results-random-mixed} and \Table{results-random-mixed-rest}.
}{%
\Table{results-random-mixed}.
}
Once again, \SrPlus{}\PPL{} showed the \emph{fastest query time on average
across all instances} for both query sets.
As the reachability in a \DAG{} is low in general (see also \Theorem{ts-facts})
and particularly in the benchmark instances, the average query times for
\Instance{random} resemble those for \Instance{negative}.
On the other hand, the results for the \Instance{mixed} query set are more
similar to those for the \Instance{positive} query set, as the relative
differences in performance among the algorithms are more pronounced there.
\Table{results-averages} compactly shows the average query time over all
instances for each query set.
Only \PPL{} and \SrPlus{}\PPL{} achieved an average query time of less than
\SI{1}{\micro\second} (and even less than \SI{0.35}{\micro\second}).

\begin{table}[tb]
\centering
\caption{%
Mean speedups with \Sreach{} plus fallback over pure fallback algorithm. Values greater \num{1.00} are highlighted.}%
\TabLabel{results-speedups-short}
\includegraphics[width=\linewidth]{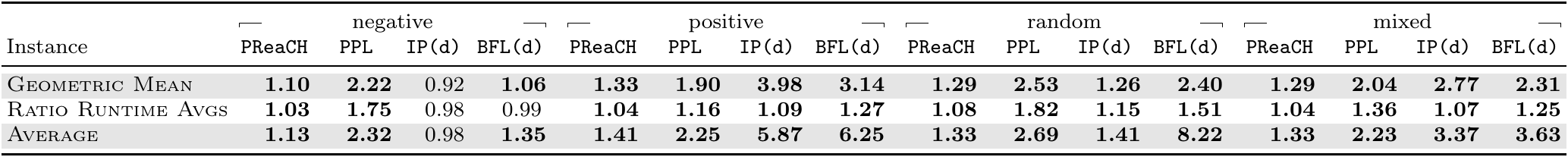}
\end{table}

\ifbool{longversion}{%
\afterpage{%
\begin{landscape}
\begin{table}[tb]
\centering
\caption{%
Speedups with \Sreach{} plus fallback over pure fallback algorithm. Values greater \num{1.00} are highlighted.}%
\TabLabel{results-speedups}
\includegraphics[width=\linewidth]{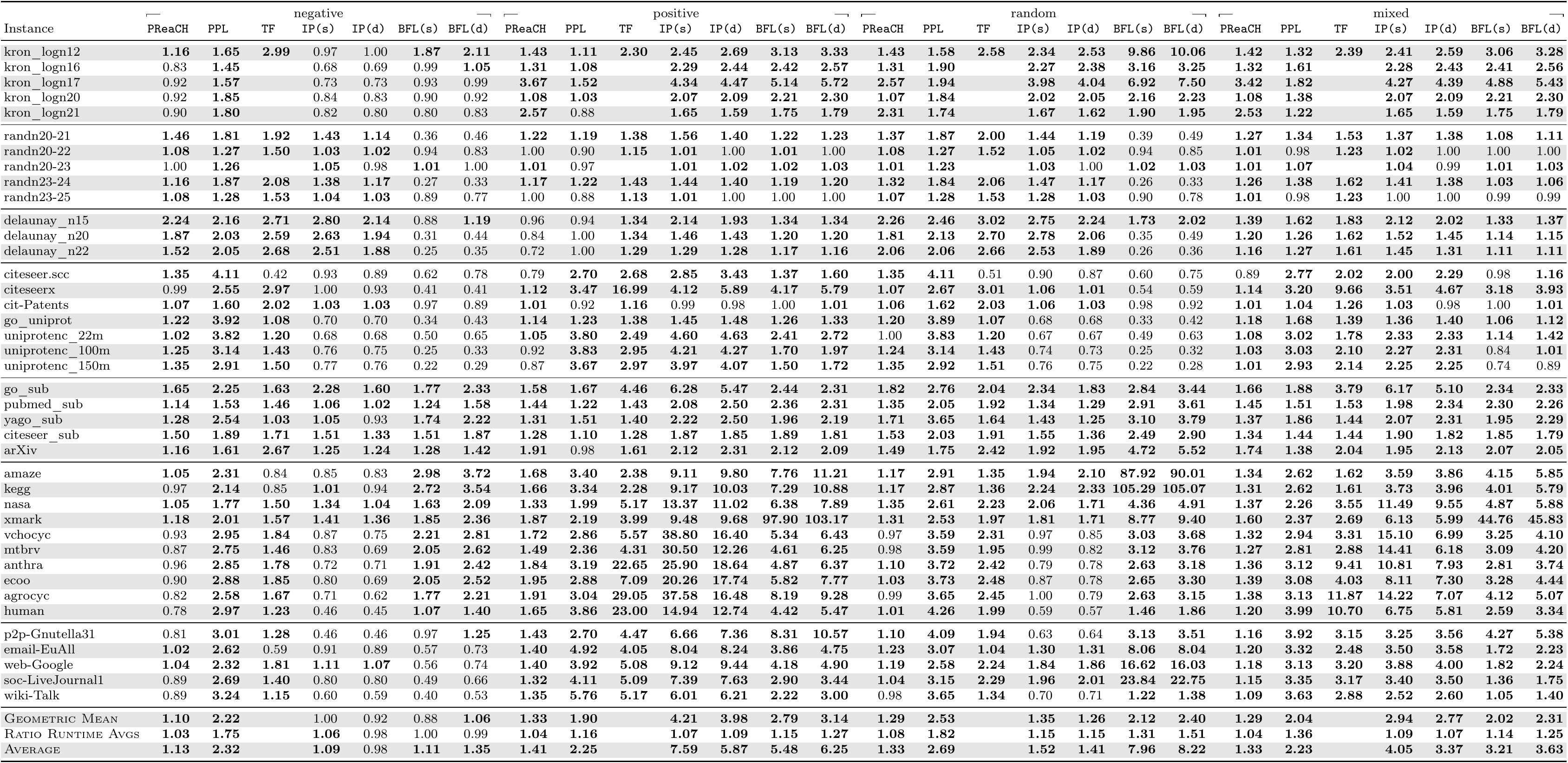}
\end{table}
\end{landscape}
}

}{}
\textbf{Speedups by \Sreach{}.}
We next investigate the relative speedup of \Sreach{} with different fallback
solutions over running only the fallback algorithms.
\Table{results-speedups} lists the ratios of the average query time of
each competitor algorithm run standalone divided by the average query time of
\Sreach{} plus that algorithm as fallback,
for all four query sets.
A compact version is also given in \Table{results-speedups-short}.
In the large majority of cases, using \Sreach{} as a preprocessor resulted in a
speedup, except in case of \Instance{negative} or \Instance{random} queries for
\BFL{} and partially \IP{} on the large real instances as well as for
\Preach{} and partially again \IP{} on the small real sparse and \SNAP{}
instances.
The largest speedup of around \num{105} could be achieved for \BFL{} on
\Instance{kegg} for random queries.
The mean speedup (geometric) is at least \num{1.29} for all fallback algorithms
on the query sets \Instance{positive}, \Instance{random}, and \Instance{mixed},
where the maximum was reached for \IPsparse{} on \Instance{positive} queries
with a factor of \num{4.21}.
Only for purely \Instance{negative} queries, \IPdense{} and \BFLsparse{} were a
bit faster alone in the mean values.
\emph{In summary}, given that the algorithms are often already faster than
single memory lookups, the speedups achieved by \Sreach{} are quite high.

\ifbool{longversion}{%
\begin{table}[htb]
\centering
\caption{
Median initialization time in \si{\milli\second} in five repetitions.
Highlighted results are the overall best.\\
As a single exception, the initialization process for \FullMatrix{} was run in
parallel.
The running time reported here corresponds to the maximum running
time of one of the \num{48} threads used and is therefore \emph{not} directly comparable
to the other running times.}%
\TabLabel{results-init}
\includegraphics[width=\linewidth]{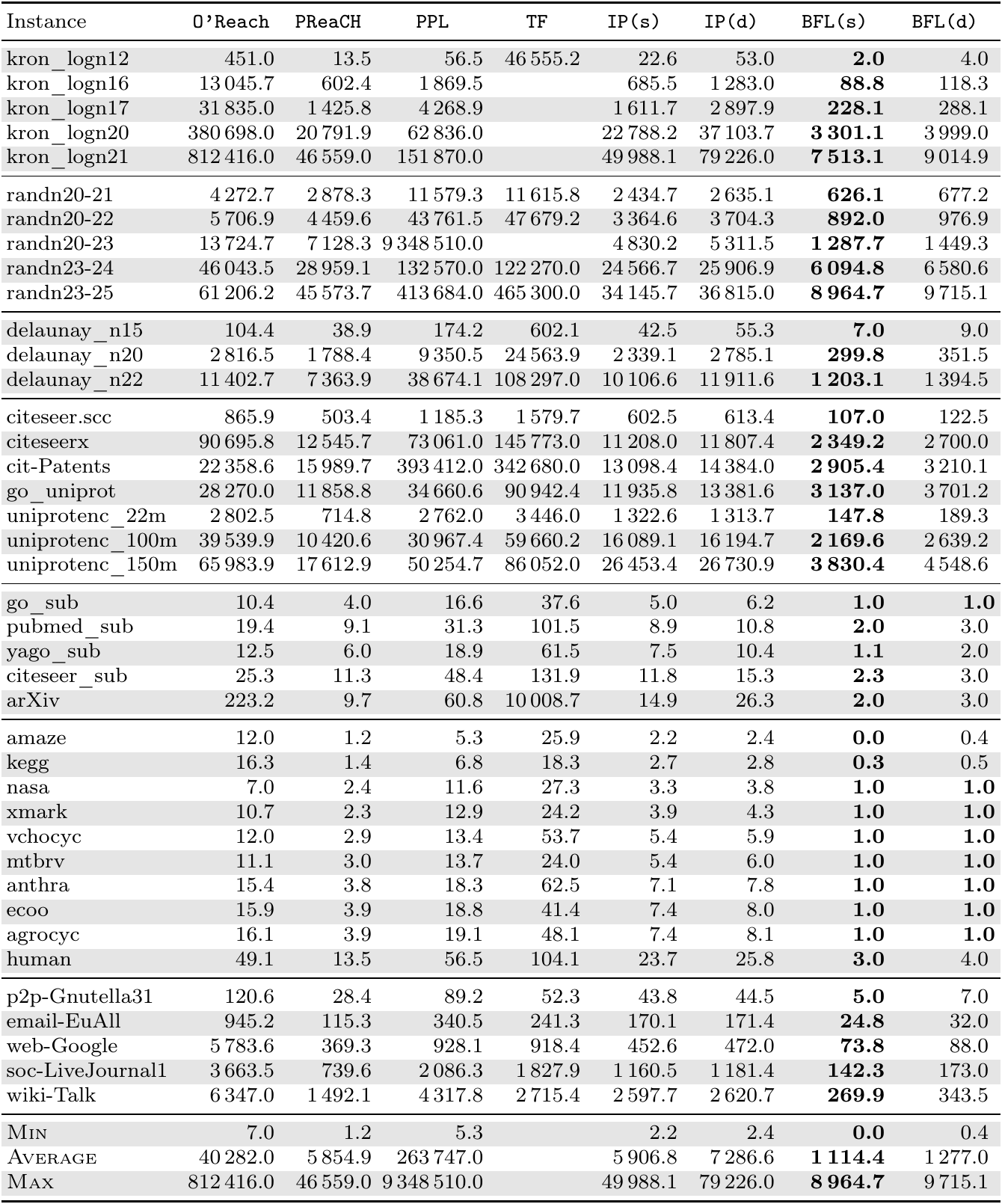}
\end{table}

\textbf{Initialization Times (\Table{results-init}).}
}{%
\textbf{Initialization Times (\Table{results-init-plus}, right).}
}
On all graphs, \BFLsparse{} had the fastest initialization time,
followed by \BFLdense{} and \Preach{}.
For \Sreach{}, the overhead of computing the comparatively large out- and
in-reachabilities of all \num{1200} candidates for $k=16$ supportive vertices
is clearly reflected in the running time on denser instances and can be reduced
greatly if lower parameters are chosen, albeit at the expense of a slightly
reduced query performance, e.g., for $k=8$.
\PPL{} often consumed a lot of time in this step, especially on denser
instances, with a maximum of \SI{2.6}{\hour} on \Instance{randn20-23}.

\textbf{Effectiveness of Observations.}
We collected a vast amount of statistical data to perform an analysis of the
effectiveness of the different observations used in \Sreach{}.

First, we look only at \emph{fast queries}, \ie, those queries that could be
answered without a fallback.
Across all query sets,
the \emph{most effective} observation was the negative basic observation on
topological orderings, \Ob{basic:topsort}, which answered around
\SI{30}{\percent} of all fast queries.
As the average reachability in the \Instance{random} query set is very low,
negative queries predominate in the overall picture.
It thus does not come as a surprise that the most effective observation is a
negative one.
On the \Instance{negative} query set, \Ob{basic:topsort} could answer
\SI{45}{\percent} of all fast queries.
After lowering the number of topological orderings to $\ParamNumTopsorts = 2$,
\Ob{basic:topsort} was still the most effective
and could answer \SI{23}{\percent} of all fast queries and \SI{33}{\percent} of
those in the \Instance{negative} query set.
The negative observations second to \Ob{basic:topsort} in effectiveness were
those looking at the forward and backward topological levels,
\Observation{basic:ts-level-fw} and~\Ob{basic:ts-level-bw}, which could
answer around \SI{15}{\percent} each on the \Instance{negative} query set
and around \SI{10}{\percent} of all fast queries.
Note that we increased the counter for \emph{all} observations that could
answer a query for this analysis, not just the first in order, which
is why there may be overlaps.
The observations using the max and min indices of extended topological
orderings, \Ob{tsp:max} and \Ob{tsp:min}, could answer \SI{9}{\percent} and
\SI{6}{\percent} of the fast queries in the \Instance{negative} query set, and
the observations based on supportive vertices, \Ob{sv:out-s-nout-t} and
\Ob{sv:nin-s-in-t}, around \SI{3}{\percent} each.
Reducing the number of topological orderings to $\ParamNumTopsorts = 2$ decreased
the effectiveness of \Ob{tsp:max} and \Ob{tsp:min} to around~\SI{5}{\percent}.

The \emph{most effective positive observation} and the second-best among all
query sets, was the supportive-vertices-based \Observation{sv:in-s-out-t},
which could answer almost \SI{16}{\percent} of all fast queries and
almost \SI{55}{\percent} in the \Instance{positive} query set.
Follow-up observations were the ones using high and low indices,
\Ob{tsp:hi} and \Ob{tsp:lo}, with \SI{18}{\percent} and \SI{16}{\percent}
effectiveness for the \Instance{positive} query set.
The remaining two, \Ob{tsp:max} and \Ob{tsp:min}, could answer \SI{6}{\percent}
and \SI{4}{\percent} in this set.
Reducing the number of topological orderings to $\ParamNumTopsorts = 2$ led to
a slight deterioration in case of \Ob{tsp:hi} and \Ob{tsp:lo} to
\SI{14}{\percent}, and to \SI{5}{\percent} and \SI{3}{\percent} in case of
\Ob{tsp:max} and \Ob{tsp:min}, each with respect to the \Instance{positive}
query set.

Among all fast queries that could be answered by \emph{only one} observation,
the most effective observation was the positive supportive-vertices-based
\Observation{sv:in-s-out-t} with over \SI{40}{\percent} for all query sets
and \SI{68}{\percent} for the \Instance{positive} query set, followed
by the negative basic observation using topological orderings, \Ob{basic:topsort},
with a bit over \SI{20}{\percent} for all query sets and \SI{52}{\percent}
for the \Instance{negative} query set.

Looking now at the entire query sets, our statistics show that
\emph{\SI{95}{\percent}} of all \emph{queries could be answered via an
observation} on the \emph{\Instance{negative}} set.
In \SI{70}{\percent} of all cases, \Ob{basic:ts-level-fw} in the second test,
which uses topological forward levels, could already answer the query.
In further \SI{16}{\percent} of all cases, the observation based on topological
backward levels, \Ob{basic:ts-level-bw}, was successful.
On the \Instance{positive} query set, the fallback rate was \SI{28}{\percent}
and hence higher than on the  \Instance{negative} query set.
\SI{52}{\percent} of all queries in this set could be answered by the
supportive-vertices-based observation \Ob{sv:in-s-out-t}, and
the high and low indices of extended topological orderings
\Ob{tsp:hi} and \Ob{tsp:lo} were responsible for another \SI{7}{\percent} each.
Observe that here, the first observation in the order that can answer a query
``wins the point'', \ie, there are no overlaps in the reported effectiveness.

\ifbool{longversion}{%
\begin{table}[tb]
\centering
\caption{Real index size in memory (in \si{\mega\byte}).}\TabLabel{memory}
\includegraphics[width=\linewidth,height=\textheight,keepaspectratio]{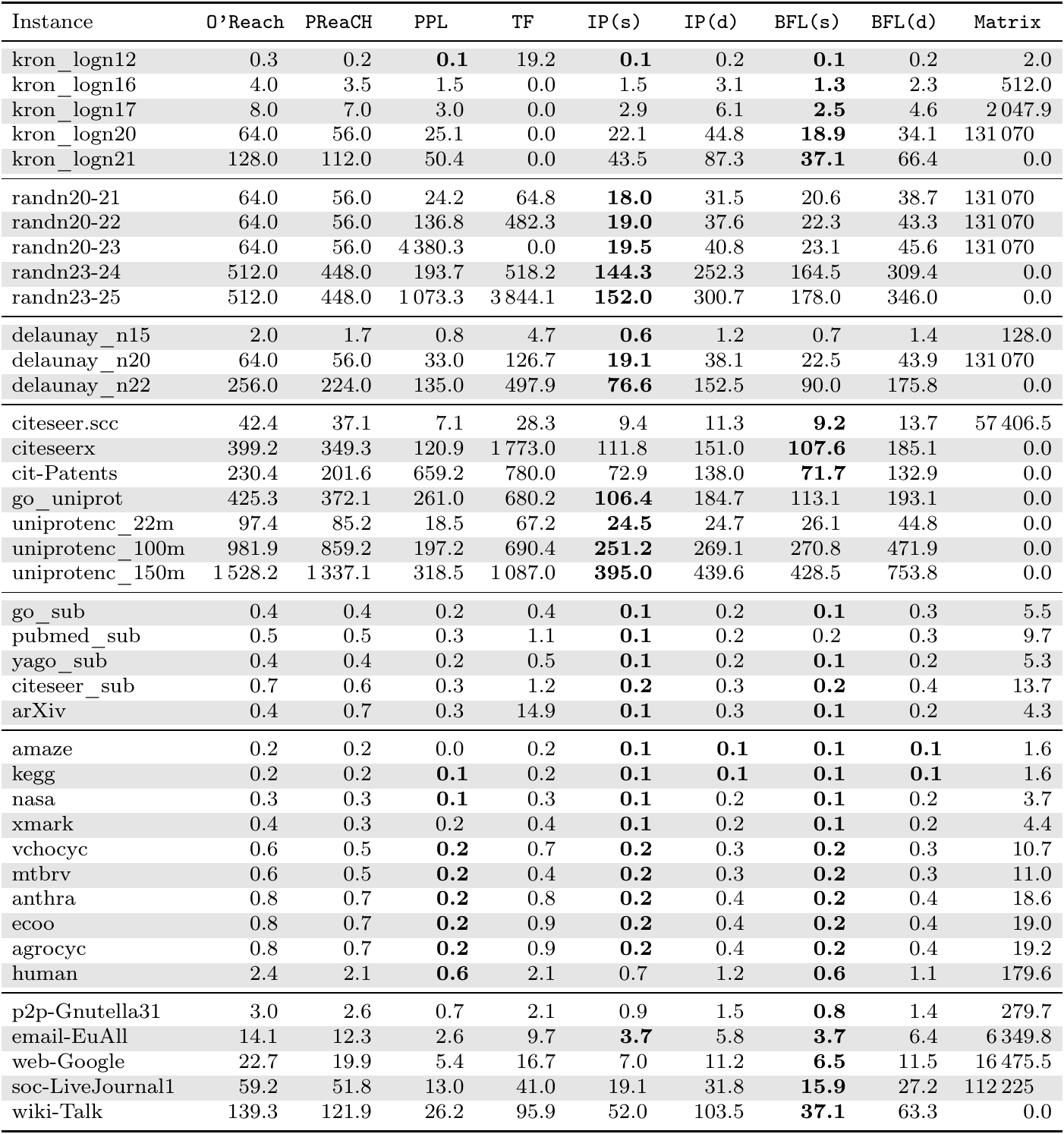}
\end{table}

}{%
}
\textbf{Memory Consumption.}
\ifbool{longversion}{%
\Table{memory} lists the memory each algorithm used for their
\emph{reachability index}.
As \Sreach{} was configured with $\ParamNumSupports = 16$ and
$\ParamNumTopsorts = 4$, its index size is $\SI[parse-numbers=false]{64n}{\Byte}$.
Consequently, the reachability indices of \Sreach{}, \Preach{}, \PPL{},
\IP{}, \BFL{}, and, with one
exception for \TF{}, fit in the L3 cache of \SI{6}{\mega\byte} for all small
real instances.
For \FullMatrix{}, this was only the case for the four smallest instances from
the small real sparse set, three of the small real dense ones, and the smallest
Kronecker graph, which is clearly reflected in its average query time
for the \Instance{negative}, \Instance{random}, and, to a slightly lesser extent,
\Instance{mixed} query sets.
Whereas for \Sreach{}, \Preach{}, %
and \FullMatrix{}, the index size
depends solely on the number of vertices, \IP{}, \BFL{}, \PPL{} and \TF{}
consumed more memory the larger the density $\frac{m}{n}$.
\IPsparse{} usually was the most space-efficient and never used more than
\SI{395}{\mega\byte}, followed by
\BFLsparse{} (\SI{429}{\mega\byte}),
\IPdense{} (\SI{440}{\mega\byte}),
\BFLdense{} (\SI{754}{\mega\byte}),
\Preach{} (\SI{1.3}{\giga\byte}),
\Sreach{} (\SI{1.5}{\giga\byte}),
and
\PPL{} (\SI{4.4}{\giga\byte}).
All these algorithms are hence suitable to handle graphs with several millions
of vertices even on hardware with relatively little memory (with respect to
current standards).
\TF{} used up to \SI{3.8}{\giga\byte}
(\Instance{randn23-25}), but required even more than \SI{64}{\giga\byte} at least
during initialization on all instances where the data is missing in the table.
}{%
Due to the configuration of \Sreach{}, its index size is
$\SI[parse-numbers=false]{64n}{\Byte}$.
Consequently, the reachability indices of \Sreach{}, \Preach{}, \PPL{},
\IP{}, \BFL{}, and, with one
exception for \TF{}, fit in the L3 cache of \SI{6}{\mega\byte} for all small
real instances.
For \FullMatrix{}, this was only the case for a few very small instances, e.g.,
from the small real sparse and real dense set,
which however is clearly reflected in the average query time there.
Whereas for \Sreach{}, \Preach{}, %
and \FullMatrix{}, the index size
depends solely on the number of vertices, \IP{}, \BFL{}, \PPL{} and \TF{}
consumed more memory the larger the density $\frac{m}{n}$.
\IPsparse{} usually was the most space-efficient and never used more than
\SI{395}{\mega\byte}, followed by
\BFLsparse{} (\SI{429}{\mega\byte}),
\IPdense{} (\SI{440}{\mega\byte}),
\BFLdense{} (\SI{754}{\mega\byte}),
\Preach{} (\SI{1.3}{\giga\byte}),
\Sreach{} (\SI{1.5}{\giga\byte}),
and
\PPL{} (\SI{4.4}{\giga\byte}).
A table listing the memory each algorithm used for their
reachability index is available in the full version~\cite{arxiv}.
}

\section{Conclusion}
In this paper, we revisited existing techniques for the static reachability
problem and combined them with new approaches to support a large portion of
\emph{reachability queries} in constant time using a linear-sized
\emph{reachability index}.
Our extensive experimental evaluation shows that
in almost all scenarios,
combining any of the existing algorithms with
our new techniques implemented
in \Sreach{} %
can speed up the query time by several factors.
In particular \emph{supportive vertices} have proven to be effective to answer
positive queries quickly.
As a further plus, \Sreach{} is flexible: memory usage, initialization time,
and expected query time can be influenced directly by three parameters, which
allow to trade space for time or initialization time for query time.
Moreover, our study demonstrates that, due to cache effects, a high investment
in space does not necessarily pay off:
\emph{Reachability queries} can often be answered even significantly faster
than single memory accesses in a precomputed full reachability matrix.

The on average fastest algorithm across all instances and types of queries was
a combination of \Sreach{} and \PPL{} with an average query time of less than
\SI{0.35}{\micro\second}.
As the initialization time of \PPL{} is relatively high, we also recommend
\Sreach{} combined with \Preach{} as a less expensive alternative solution with
respect to initialization time and partially also memory, which still achieved
an average query time of at most \SI{11.1}{\micro\second} on all query sets.

\ifbool{longversion}{
}{%
\clearpage

}
\clearpage

\bibliographystyle{plainurl}
\bibliography{paper}
\end{document}